\documentclass[useAMS,usenatbib]{mn2e}
\usepackage[activeacute,english]{babel}
\usepackage[latin1]{inputenc}
\usepackage[dvips]{graphicx,color}
\usepackage{graphicx}
\usepackage[T1]{fontenc} 
\usepackage{aecompl} 
\usepackage{epsfig}
\usepackage{latexsym}
\usepackage{amssymb}
\usepackage{amsmath}
\usepackage{lscape}
\usepackage{rotating}
\usepackage{verbatim}
\usepackage{natbib}
\usepackage{caption}
\usepackage{multirow}
\usepackage{float}

\usepackage{soul}

\IfFileExists{.pdfmode}{%
  \def\UseColor{true}
  \usepackage{ae,aecompl} 
  }{
  \def\UseColor{false}
  }

\usepackage[pdftitle={Tracking Cluster Debris (TraCD) I: dissolution of clusters and searching for the Solar cradle},%
pdfauthor={Guido R. I. Moyano Loyola},%
pdfsubject={},%
pdfkeywords={},%
colorlinks=\UseColor,%
dvips%
]{hyperref}

\title[Tracking Cluster Debris (TraCD) I]
  {Tracking Cluster Debris (TraCD) - I. Dissolution of clusters and searching for the solar cradle}
\author[G. R. I. Moyano Loyola et al]
  {Guido R. I.~Moyano Loyola$^1$\thanks{Email: gmoyano@astro.swin.edu.au},
    Chris Flynn$^1$,
    Jarrod R.~Hurley$^1$, 
    Brad K. Gibson$^{2,3}$\\
  $^1$Centre for Astrophysics and Supercomputing, Swinburne University of Technology, PO Box 218, Hawthorn, VIC 3122, Australia\\
  $^2$E.A. Milne Centre for Astrophysics, University of Hull, Hull HU6 7RX, UK\\
  $^3$Jeremiah Horrocks Institute, University of Central Lancashire, Preston PR1 2HE, UK}
\date{Released 2014 Xxxxx XX}

\pagerange{\pageref{firstpage}--\pageref{lastpage}} \pubyear{2014}

\def\LaTeX{L\kern-.36em\raise.3ex\hbox{a}\kern-.15em
    T\kern-.1667em\lower.7ex\hbox{E}\kern-.125emX}

\begin{document}

\label{firstpage}

\maketitle

\begin{abstract}
The capability to reconstruct dissolved stellar systems in dynamical and chemical space is a key factor in improving our understanding of the evolution of the Milky Way. Here we concentrate on the dynamical aspect and given that a significant portion of the stars in the Milky Way have been born in stellar associations or clusters that have lived a few Myr up to several Gyr, we further restrict our attention to the evolution of star clusters. We have carried out our simulations in two steps: (1) we create a simulation of dissolution and mixing processes which yields a close fit to the present-day Milky Way dynamics and (2) we have evolved three sets of stellar clusters with masses of 400, 1\,000 and 15\,000 M$_{\odot}$ to dissolution. The birth location of these sets was 4, 6, 8 and 10 kpc for the 400 and 1\,000 M$_{\odot}$ clusters and 4, 6, 8, 10 and 12 kpc for the 15\,000 M$_{\odot}$. We have focused our efforts on studying the state of the escapers from these clusters after 4.5 Gyr of evolution with particular attention to stars that reach the Solar annulus, i.e. 7.5 $\le$ R$_{\rm{gc}}$ $\le$ 8.5 kpc.
We give results for Solar twins and siblings over a wide range of radii and cluster  masses for two dissolution mechanisms. From kinematics alone, we conclude that the Sun was $\sim$50 per cent more likely to have been born near its current Galactocentric radius, rather than have migrated (radially) $\sim$ 2 kpc since birth. We conclude our analysis by calculating magnitudes and colours of our single stars for comparison with	the samples that the {\it{Gaia}}, {\it{Gaia}}-ESO and GALAH-AAO surveys will obtain. In terms of reconstructing dissolved star clusters we find that on short time-scales we cannot rely on kinematic evolution alone and thus it will be necessary to extend our study to include information on chemical space. 
\end{abstract}

\begin{keywords}
sun: general \ -- stars: kinematics and dynamics \ -- galaxy: kinematics and dynamics \ -- open clusters and associations: general \ -- solar neighbourhood.
\end{keywords}

\section{Introduction}
\label{Intro}

Reconstructing the history of the Milky Way from the kinematics and chemistry of its stellar components is one of the grand challenges of modern galactic astrophysics \citep{Freeman}, as we enter an era of huge automated surveys of many millions of stars, e.g. {\it{Gaia}} \citep{Perryman}.

In this work, we construct models of dissolving star clusters and use them to analyse the
internal stellar assembly history of the disc of our Galaxy. Of particular importance is to identify the extent to which it is possible to reconstruct dissolved stellar clusters and information that can be shed on the possible parent cluster of the Sun.

Recent steps in tracking stellar clusters within kinematic space have been taken by \citet{FujiiBaba}, who  analysed the destruction of star clusters by radial migration using the {\tt{BRIDGE}} code \citep{Fujii}. The authors integrated a Milky Way-like disc up to an age of 5 Gyr and when the self-excited spiral arms
developed, i.e. overdensities on the disc, star clusters were placed `by hand' on those overdense regions.

We face this challenge from an alternative, yet complementary perspective. We extend the limits of
{\tt{NBODY6}} -- the state-of-the-art code for following the internal evolution of star clusters -- by post-analysing the stars from dissolved clusters with an orbit integrator and at a future stage a Galactic chemodynamical code. In this way, we obtain accurate predictions for the dynamical evolution of stellar clusters and the subsequent kinematics of stars that escape from these clusters into the Galactic disc, helping to constrain efforts to unravel the chemical evolution history of the Galaxy \citep{Freeman}. 

We have studied the distribution in velocity space of the stars escaping from star clusters within the
Milky Way \citep{Guido}. In that work, we could identify the mechanisms that produce
escapers within different velocity ranges and determine the maximum mass with which a given escaper can
reach the Galactic halo as a main-sequence or giant star.
The next step is to track the evolution of previously escaping stars as they populate the Galactic disc and to subsequently test the limits to which they can be tracked to a common origin.

It is expected that the diffusive nature of the disc will eventually be effective in erasing the kinematic
initial conditions of stars and thus the dynamical information of dissolved clusters will not be sufficient
for a complete reconstruction. We make estimates here of how long this takes for a variety of cluster masses and Galactocentric radii.

Increasing evidence (\citealt{Tutukov}; \citealt{Carpenter}; \citealt{Lada}; \citealt{Porras}; \citealt{Bressert}; \citealt{Kruijssen}) suggests that most stars are born in associations of some sort. The Sun seems to not be an exception (\citealt{Gaidos1995}; \citealt{Adams}; \citealt{Pichardo}; \citealt{Pfalzner}). Extensive work suggests that the Sun has migrated outwards 1$-$3 kpc since its birth, based upon a wide range of (primarily) chemical constraints (\citealt{Nieva} and references therein). Finding the potential parent cluster of the Sun and locating solar siblings, i.e. stars that were born with the Sun, can place constraints on the dynamical and chemical evolution of the Galactic disc in the last $\sim$ 4.5 Gyr.

We have carried out our simulations in three stages. In Section \ref{STRUCTURE OF THE DISC}, we show the first stage, which involves how the background sea of stars was created using an orbit integrator which simulates the effects of orbital diffusion (presumably due to asymmetries) in the Galactic potential. We aim to make this background sea as comparable with stellar density and kinematic observations as possible \citep{Bovy}.
In Section \ref{EVOLUTION OF SMALL OPEN CLUSTERS WITHIN THE GALACTIC DISC}, we describe the second stage where we evolved small star clusters until dissolution with {\tt{NBODY6}} \citep{Aarseth3} under the effect of a time-independent potential of the Galaxy. After escaping, these stars are fed into the background sea of stars and integrated in a time-dependent Milky Way potential. Section \ref{The search for the Sun} is devoted to analysing whether the Sun could be born from any of our modelled clusters and how our models compare with upcoming surveys such as {\it{Gaia}} \citep{Perryman}. We summarize our results and conclude in Section \ref{SUMMARY}.

\section{THE BACKGROUND SEA OF STARS}
\label{STRUCTURE OF THE DISC}

Most disc-like galaxies can be characterized as a highly flattened structure with an exponential radial scalelength but below sub-kpc scales the picture is no longer so simple, e.g. presence of spiral arms and bars. Refinements of the treatment of galactic potentials, i.e. moving from point-mass models to full three-dimensional potentials that include features like spiral arms and three-dimensional velocity ellipsoids, have been carried out for more than two decades (e.g. \citealt{SB02}). However, in a few years {\it{Gaia}} \citep{Perryman} will give us an even better picture of the Galaxy for constraining models.

It is now well established that radial mixing of stellar orbits is ubiquitous in disc galaxies (e.g. simulations: \citealt{SB02}; \citealt{Sanchez-Courty}; \citealt{Grand}; observations: \citealt{Bird}; \citealt{Haywood}; \citealt{Yu}). There are four main mechanisms that are proposed to produce radial mixing: interaction with giant molecular clouds (GMCs; \citealt{Spitzer}), interaction with transient spiral arms \citep{Barbanis}, massive compact halo object impacting the disc \citep{Lacey} and satellite infall \citep{Velazquez}. 

One of the more evident mechanisms that changes the orbits of stars are the spiral arms, although whether these are fixed or transient features of discs is a matter of discussion (\citealt{Lindblad}; \citealt{LinShu}; \citealt{Goldreich}; \citealt{Athanassoula}; \citealt{FujiiBaba}).

We focus in this paper on two of the mechanisms: mixing through interactions with GMCs and with  transient spiral arms. \citet{SB02} introduced a new terminology for these two effects, `blurring' for when the epicycle amplitude of the orbit changes but there is little change in the angular momentum owing to a scattering event with a GMC, and `churning' when the guiding-centre of the orbit changes without changing the angular momentum owing to interactions with transient spiral arms.

In the remainder of this section, we describe our model for the Galactic potential, how we populate the disc with stars, how we model stellar blurring and churning and how we constrain this population kinematically through comparison with observations. The aim is to use a disc composed of kinematic mixed galactic populations as the background sea from where we will aim to identify cluster debris.

\subsection{Orbit Integrator -- Galactic disc}
\label{disc-model}

\subsubsection{Orbit Integrator}
\label{integrator}

We have used our own code {\tt{GALORB}} ({\tt{GO}} hereafter) which integrates orbits by using a a standard Runge-Kutta \citep{NR} method for integrating the equations of motion with an adaptive timestep. This code runs on a single-core CPU. The treatment of the Galactic potential can be divided into a static component and techniques to model time-dependent features.

First the static potential that this integrator uses is represented by a three-component Galaxy: bulge, disc and halo. A summary of the different component parameters can be seen in Table \ref{pot-table}.

\begin{table*}
 \caption{Properties of the Galactic model used in Section \ref{integrator} and in {\tt{NBODY6}} (see Section \ref{EVOLUTION OF SMALL OPEN CLUSTERS WITHIN THE GALACTIC DISC}). The orbit integrator ({\tt{GO}}) uses two bulges, i.e. spheroid and central components, represented by Plummer spheres. {\tt{NBODY6}} uses a point mass bulge. Both the orbit integrator and {\tt{NBODY6}} uses \citet{Miyamoto} discs and logarithmic haloes, although the orbit integrator uses a combination of three discs.}
 \label{pot-table}
 \begin{tabular}{@{}lrrrrr}
  \hline
  Component & Mass & Scalelength  & Scaleheight  & Core radius & Circular velocity\\ 
  & ($\times 10^{10}$ M$_{\odot}$) & (kpc) & (kpc) & (kpc) & (km s$^{-1}$)\\
  \hline
  Bulge - spheroid ({\tt{GO}}) & 0.03 & 2.7 & \--- & \--- & \--- \\
  Bulge - central ({\tt{GO}}) & 0.6  & 0.42 &\--- &\--- &\---\\
  Bulge pointmass ({\tt{NBODY6}}) & 0.6 &\--- &\--- &\--- &\---\\
  Disc MN-1 ({\tt{GO}})    & 5.3 & 2.9 & 0.3 & \--- &\---\\
  Disc MN-2 ({\tt{GO}})    & $-$2.3 & 8.7 & 0.3 & \--- &\---\\
  Disc MN-3 ({\tt{GO}})    & 0.26 & 17.4 & 0.3 & \---&\---\\
  Disc MN ({\tt{NBODY6}}) & 6 & 2.1 & 0.3 &\--- &\---\\
  Logarithmic Halo ({\tt{GO}}/{\tt{NBODY6}}) &\--- &\--- &\--- & 8 &220 (at 8 kpc)\\
  \hline
 \end{tabular}
\end{table*}

The bulge is composed of a spheroid component represented as a Plummer sphere with a mass of 0.3 $\times$ 10$^{9}$ M$_{\odot}$ and a scale of 2.7 kpc, plus a central component represented as a Plummer sphere with a mass of 6.0 $\times$ 10$^{9}$ M$_{\odot}$ and a scale of 0.42 kpc

Three \cite{Miyamoto} discs have been employed with the following parameters: (a) mass of 5.3 $\times$ 10$^{10}$ M$_{\odot}$ with a radial scalelength of 2.9 kpc, (b) mass of $-2.3$ $\times$ 10$^{10}$ M$_{\odot}$ with a radial scalelength of 8.7 kpc and (c) mass of 2.6 $\times$ 10$^{9}$ M$_{\odot}$ with a radial scalelength of 17.4 kpc \citep{Sommer-Larsen}. All discs have a vertical scaleheight of 0.3 kpc (which  gives a volume density of 0.05 M$_{\odot}$ pc$^{-3}$ at 8 kpc). These three discs together give an exponential surface density profile over a large range of Galactocentric radius. We have chosen to use Miyamoto discs because it is simple to implement them in the orbit integrator, albeit one of them with negative mass, to fine-tune the fit to an exponential profile.

Finally, we set a logarithmic halo by requiring a circular velocity of 220 km s$^{-1}$ at a distance of 8.0 kpc. Our halo has a core $r_{0}$ = 8 kpc.

\begin{figure}
\includegraphics[width=9.4cm]{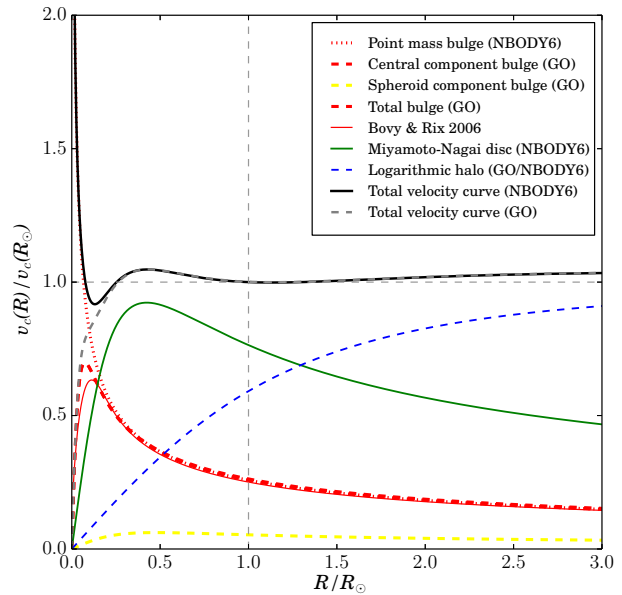}
 \caption{Circular velocity curves for the different components of the Galaxy models used in this work, with special attention to the bulge component. We have compared different bulge models: point-mass, Plummer sphere and Hernquist models \citep{Bovy}. Since the contribution of the spheroid component of the {\tt{GO}} bulge is small, both the {\tt{GO}} central and total bulge overlap. We also show the total velocity (black curve) created by combining a point-mass bulge, a single Miyamoto \& Nagai disc and a logarithmic halo. The potential that generated this curve will be used for the setup in {\tt{NBODY6}} (see Section \ref{EVOLUTION OF SMALL OPEN CLUSTERS WITHIN THE GALACTIC DISC}). This is compared to the total velocity (grey dashed curve) for the {\tt{GO}} orbit integrator. It is evident that both total velocity curves are similar for the regions that we will focus on in this work (R$_{\rm{gc}} \ge 3$ kpc). The vertical dashed line is the normalization on distance by using $R_{\rm{gc},\odot}=8$ kpc. The horizontal dashed line is the normalization for the circular velocity at $R_{\odot}$ which was set equal to $220$ km s$^{-1}$.}
 \label{velcurve}
\end{figure}

It is uncertain whether giant carbon-oxygen molecular clouds within 4 kpc of the Galactic Centre follow the total velocity curve of the Galaxy, where an increase of the circular velocity curve of these clouds has been observed (e.g. \citealt{Clemens}; \citealt{Sofue}; \citealt{Bobylev}; \citealt{Bhattar}), or if their kinematics are driven by local anisotropies (i.e. the Galactic bar). In this work, we have decided to fine-tune our Galactic model to match the latest results from \citet{Bovy}, in which they have found unprecedented constraints of the Galactic potential by using SEGUE data \citep{Yanny09}.

In Fig. \ref{velcurve}, we have plotted the rotation velocity curves for the different components of our model Galaxy with a particular focus on the bulge. The inner part of the Milky Way is effectively a whirlpool in which the dynamics will be dominated by the Galactic bar. Modelling the Galactic bar is outside the scope of this work. The combined velocity curve for our model Galaxy is shown in Fig. \ref{velcurve}.

As mentioned earlier, we aimed to mimic two key time dependent features for our analysis: (1) {\emph{Molecular clouds}}, diffuse stars by giving them a kick in velocity space chosen randomly from an isotropic Gaussian distribution with $\sigma$ = 35 km s$^{-1}$ and (2) {\emph{Spiral Arms}}, stars are churned (i.e. shifted) in radius \citep{SB02} by a random Gaussian with $\sigma$ = 20 pc around $\mu$ = 0 pc, i.e. no net shift. Both of these effects were imposed every timestep. These values of 35 km s$^{-1}$ and 20 pc have been chosen by making initial guesses and running the model, and then adjusting to achieve consistency with the observations of \citet{Lewis} as described in detail in Section \ref{Modelled velocity dispersion and observations}.

Stars are affected by both processes, i.e. diffusion and churning, every timestep.

\subsubsection{Galactic disc}
\label{galdisc}

\begin{figure}
\includegraphics[width=8.4cm]{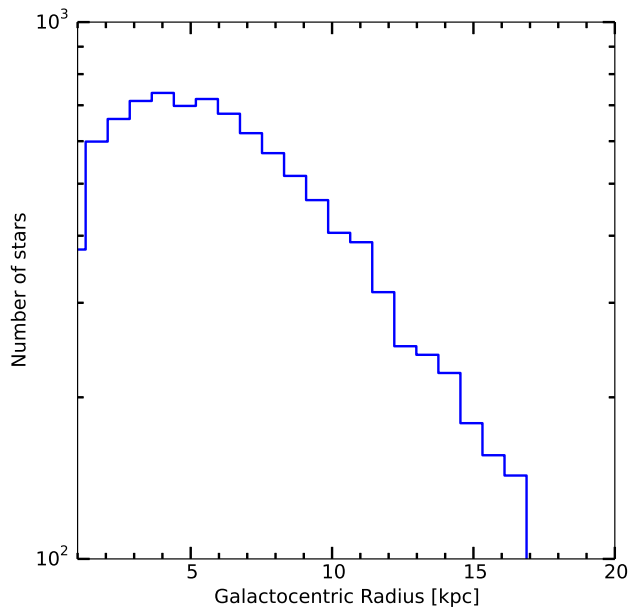}
\caption{Initial radial distribution of the stars in our disc showing number of stars in 1 kpc annuli. The stars are distributed exponentially, with a scalelength of 4 kpc.}
\label{disc}
\end{figure}

For the initial conditions of our Galactic disc we have set an exponential distribution \citep{BT08} for the position of 10\,000 stars (see Fig.  \ref{disc}). The radial and vertical velocities of these stars were taken from a Gaussian distribution with $\mu$ = 0 km s$^{-1}$ and $\sigma$ = 5 km s$^{-1}$, while the tangential velocity was also taken from a Gaussian distribution but with $\mu$ set to the local circular velocity in km s$^{-1}$ (the `local circular velocity' was taken from the total {\tt{GO}} potential of our model Milky Way presented in Fig. \ref{velcurve}) and $\sigma$ = 5 km s$^{-1}$. We have chosen the value of the  dispersion ($\sigma$) to match the average dispersion \citep{Wilson} of the interstellar medium.

\begin{figure*}
\includegraphics[width=17.5cm]{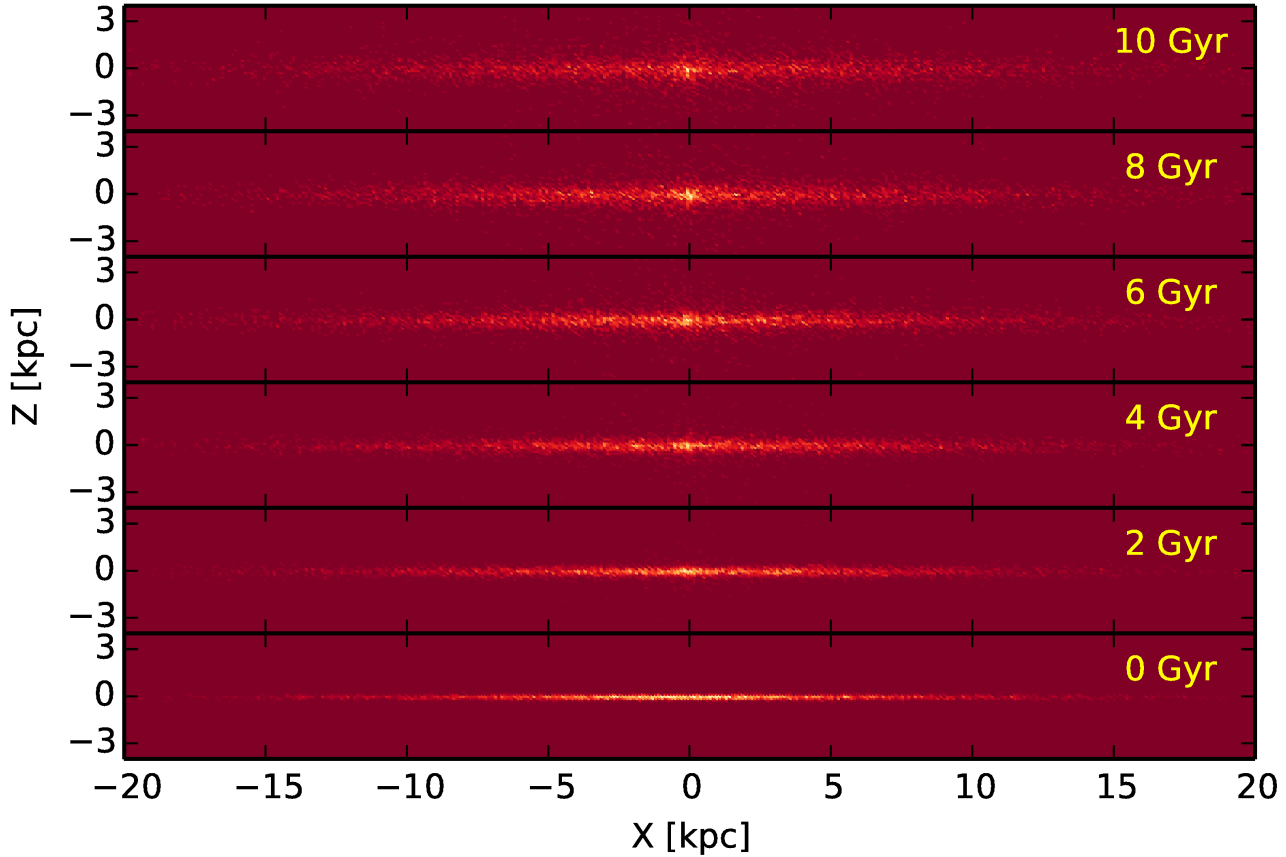}
 \caption{Evolution of the exponential disc integrated with {\tt{GO}} over 10 Gyr, shown every 2 Gyr in the ($X$,$Z$) plane. The modelled disc is a good match to a range of present day disc properties, both locally and over a large Galactocentric radius (see Section \ref{Modelled velocity dispersion and observations}).}
 \label{stack}
\end{figure*}

We have evolved the disc for 10 Gyr with the orbit integrator presented in the previous section. 
In order to have better mixed kinematic populations in the disc, i.e. old and young population of stars, we have stacked snapshots of the disc every gigayear. The final results is a disc composed of 110\,000 stars to  which we will add stellar clusters at different stages of dissolution to examine the extent to which we are able to identify them (see Sections \ref{EVOLUTION OF SMALL OPEN CLUSTERS WITHIN THE GALACTIC DISC} and \ref{The search for the Sun}). In Fig. \ref{stack}, we show how the disc heats in both the vertical and horizontal directions as it evolves. The disc clearly thickens as a function of time and is a good match to the local age-velocity relation (AVR: see the next section).

\subsection{Modelled velocity dispersion and observations}
\label{Modelled velocity dispersion and observations}

We have converted the positions and velocities of our stars to an inertial frame with the origin at the Galactic Centre and the $X$-axis pointing towards the Sun, the $Y$-axis at 90 degrees in a direction looking from the Galactic North Pole and finally the $Z$-axis pointing to the Galactic North Pole. Special care was taken in changing these new phase-space coordinates into radial and tangential components in order to compare with observations.

As noted in Section \ref{disc-model}, we have churned and blurred our Galactic disc for 10 Gyr with the described prescriptions and as a check we now compare our results to those of the Geneva-Copenhagen survey (GCS; \citealt{Holmberg}) and the work of \citet{Lewis}.

\subsubsection{Velocity Dispersion versus Age}
\label{HolmbergS}

The largest and most comprehensive stellar chemistry and kinematic survey/compilation started in the early 2000's was the Geneva-Copenhagen Survey (GCS: \citealt{Nordstrom}; \citealt{Holmberg}). This survey provides the best pre-{\it{Gaia}} determinations of ages, distances, metallicities and kinematics of nearby stars.

One of the most remarkable achievements of the GCS survey was to accurately measure the slope of the AVR in all three components, i.e $U$, $V$, $W$, highlighting that the velocity dispersion of stars increases in time. This heating is mainly a consequence of the dynamical interaction of  stars with GMCs and spiral perturbations and to a minor degree infalling satellite galaxies.

\begin{figure}
\includegraphics[width=8.4cm]{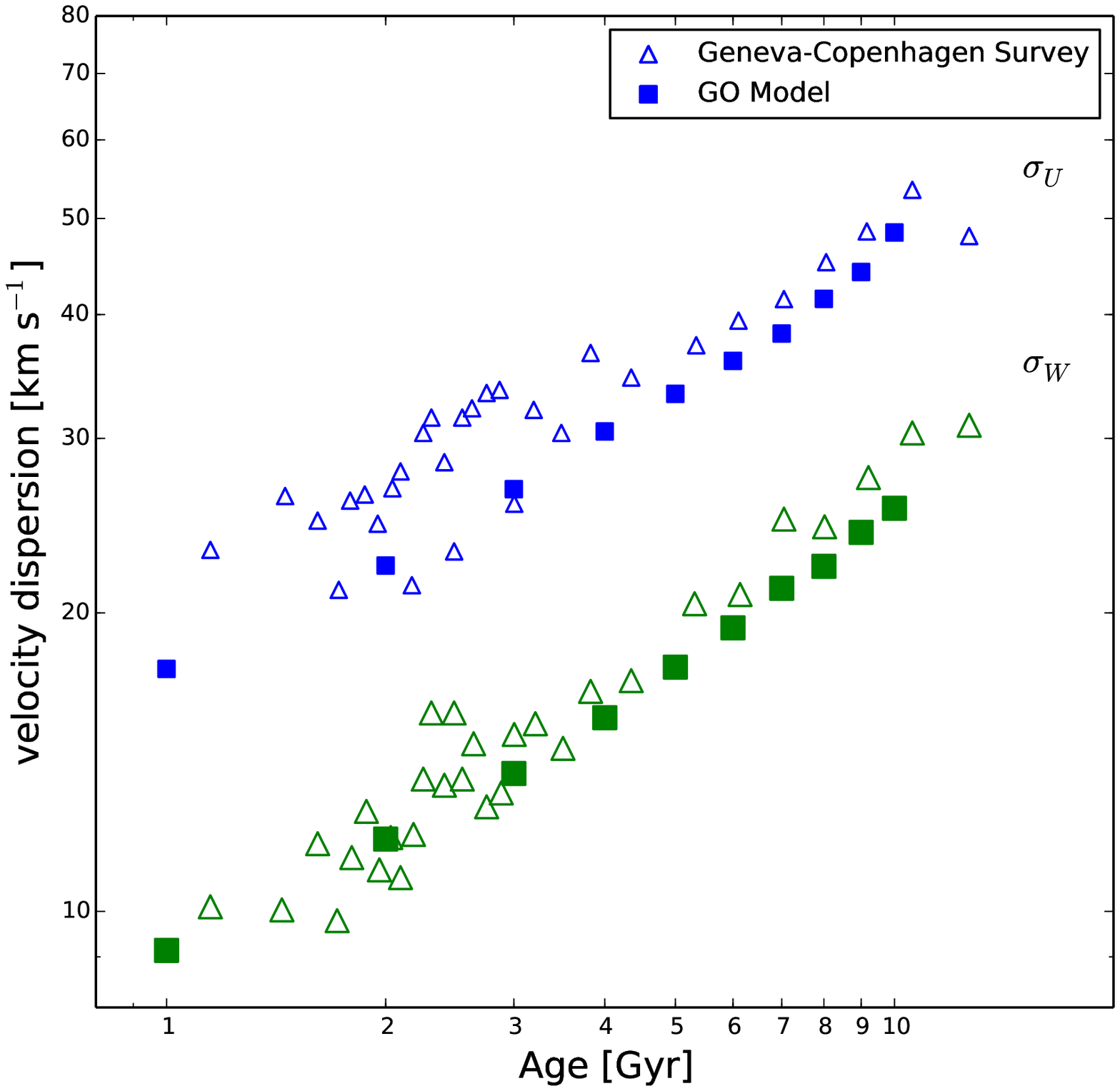}
 \caption{Comparison between the age versus velocity dispersion relation from GCS \citep{Holmberg} and our modelled Galactic disc for the velocity dispersion in the $U$-direction ($\sigma_{U}$) and the dispersion in the $W$-direction ($\sigma_{W}$). For this analysis we have taken stars in a 1 kpc annulus around the Sun (7.5 $\le$ $\rm{R}_{\rm{gc}}$ $\le$ 8.5 kpc)}
 \label{HolmbergU}
\end{figure}

\begin{figure}
\includegraphics[width=8.4cm]{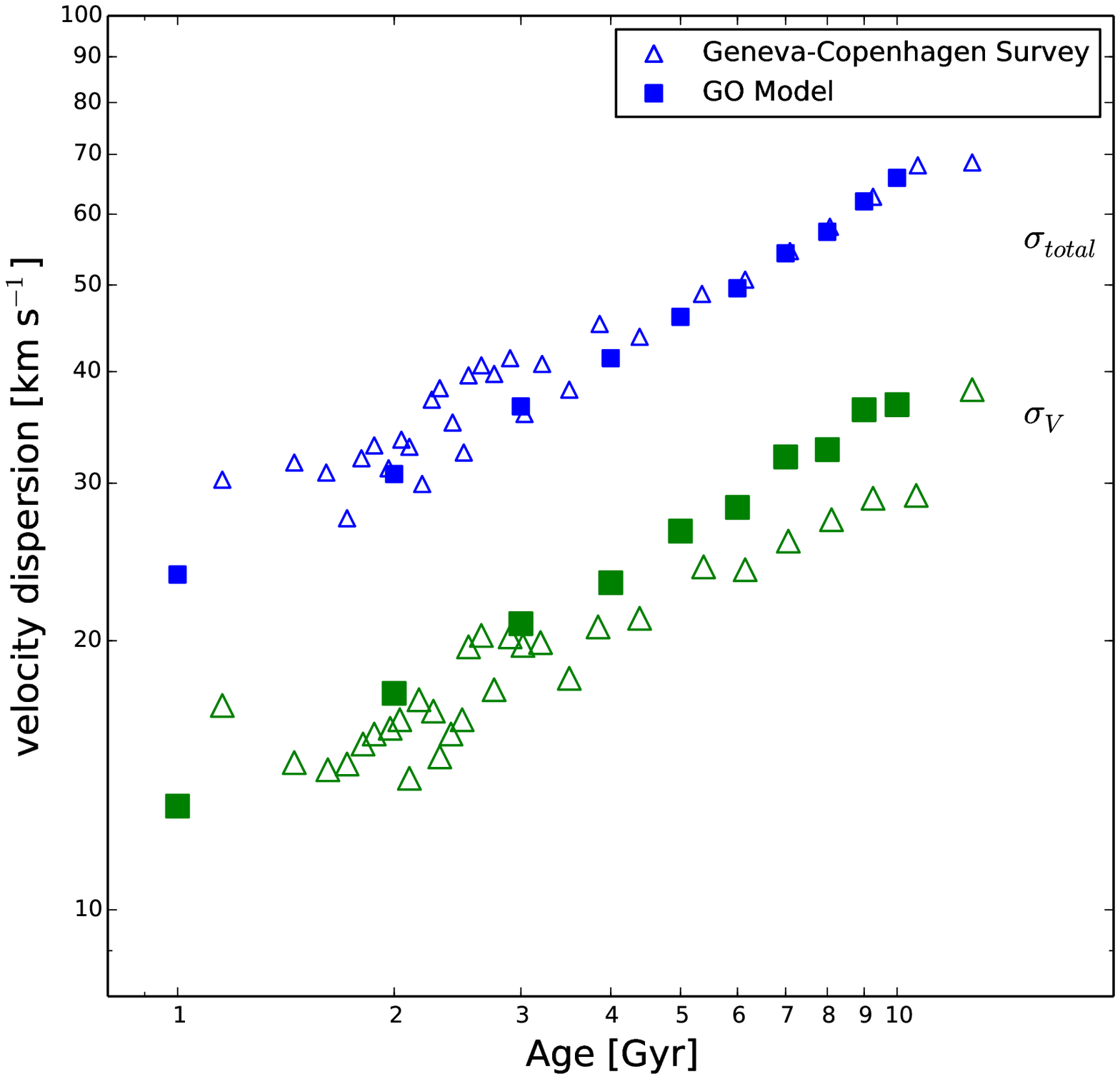}
 \caption{Same as Fig. \ref{HolmbergU} but for the total velocity dispersion ($\sigma_{total}$) and the velocity dispersion on the $V$-direction ($\sigma_{V}$).}
 \label{HolmbergV}
\end{figure}

We have considered these relations a benchmark in our analysis, and our modelled disc is a close match to the observational data, as shown in Figs. \ref{HolmbergU} and \ref{HolmbergV}. To make a consistent comparison, we have taken stars within a 1 kpc annulus from the Sun (7.5 kpc $\le$ R$_{\rm{gc}}$ $\le$ 8.5 kpc) and we have tuned our parameters to match the kinematic data from the GCS survey.

\subsubsection{Velocity Dispersion versus Radius}
\label{LewisS}

An early survey of the velocity dispersion of stars in the Galactic disc over a range of Galactocentric distances was made by \citet{Lewis}. They found that the radial and tangential velocity dispersions of disc stars fall exponentially outwards, with twice the scalelength of the visible stars. %
This result can be explained as a consequence of the higher density of molecular clouds in the inner parts of the Galaxy, which will more effectively heat stars on those regions. 

We model this by including a radial dependence of the diffusion,

\begin{eqnarray}\label{diffusion}
         \sigma_{i} & \propto & exp(-(\rm{R}-\rm{R}_{\rm{gc}})/2h_{\rm{R}}),
\end{eqnarray}

\noindent where R$_{\rm{gc}}$ = 8 kpc, $2h_{\rm{R}}$ = 8 kpc and $i = X, Y, Z$.

\begin{figure}
\includegraphics[width=8.4cm]{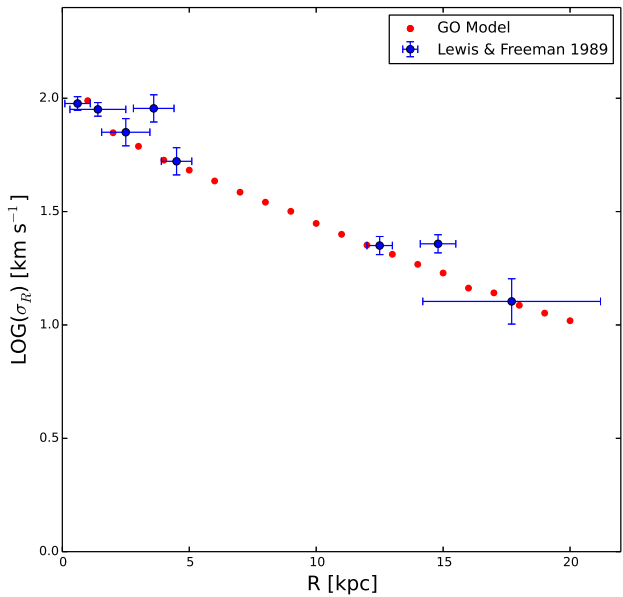}
 \caption{Comparison of the radius versus velocity dispersion relation from \citet{Lewis} and our modelled Galactic disc for the velocity dispersion in the radial direction ($\sigma_{R}$).}
 \label{LewisU}
\end{figure}

\begin{figure}
\includegraphics[width=8.4cm]{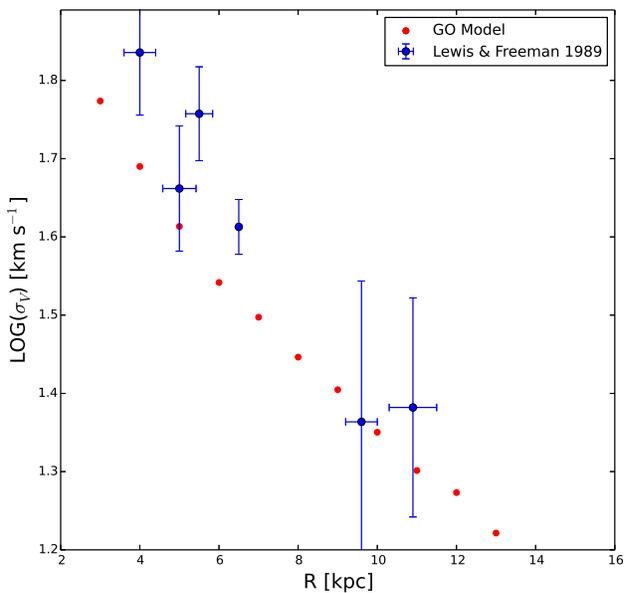}
 \caption{Same as Fig. \ref{LewisU} but for the (tangential) velocity dispersion $\sigma_{V}$.}
 \label{LewisV}
\end{figure}

Once this dependence was implemented, we obtained a close match between our models and the \citet{Lewis} 's observational data for the radial velocity dispersion out to 20 kpc as shown in Fig. \ref{LewisU}. In Fig. \ref{LewisV} we compare the tangential velocity dispersion showing that the model follows nicely the observations in the external parts of the Galaxy but underestimates the velocity dispersion of the central regions.

\subsection{Radial Migration}

As mentioned in Section \ref{STRUCTURE OF THE DISC}, several mechanisms can produce a gradual heating of the Galactic disc. In this work we have explored different prescriptions for two of the main drivers of migration, e.g. diffusion and churning. 

\begin{figure*}
\centering
\begin{tabular}{cc}
\includegraphics[width=0.45\textwidth]{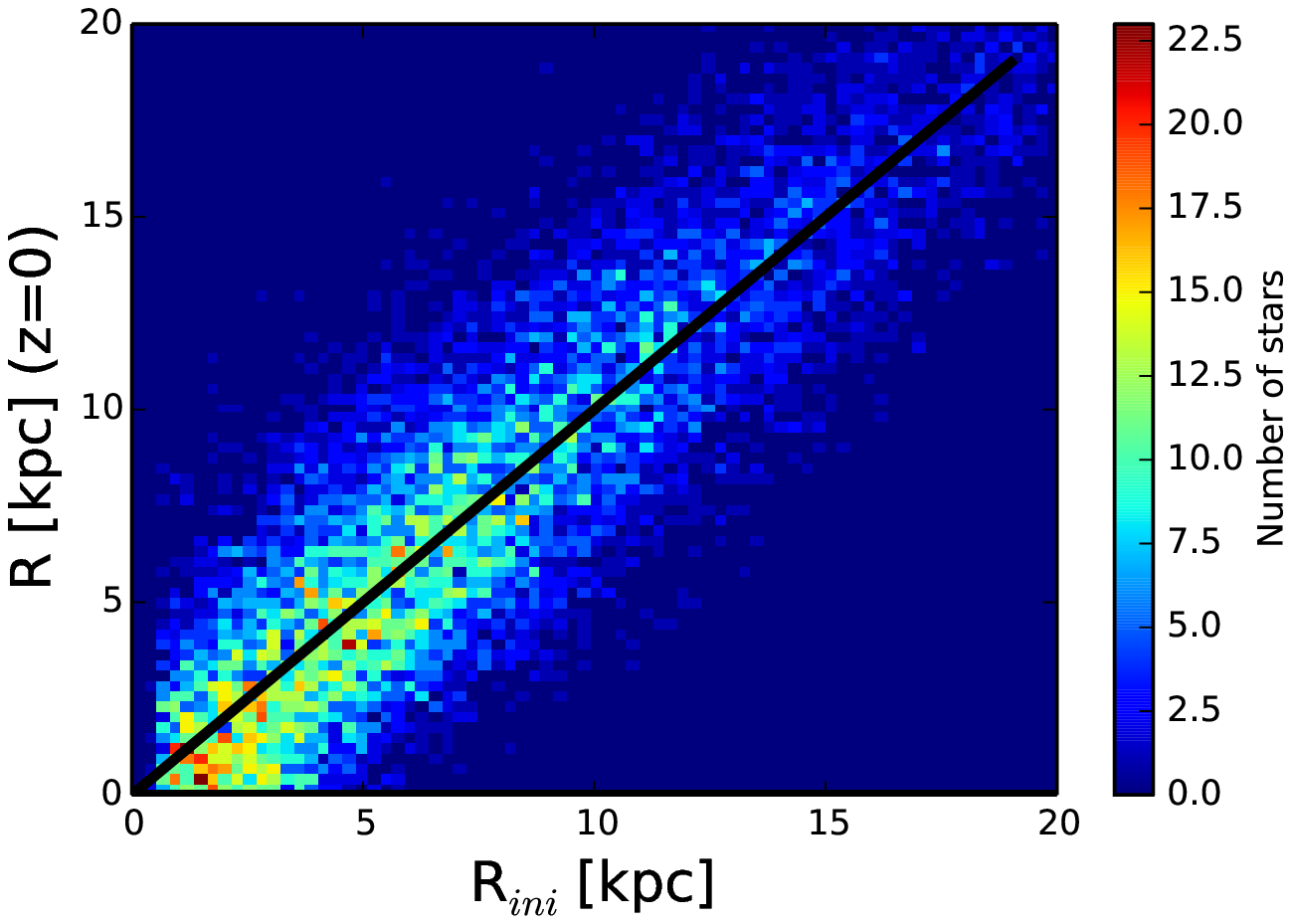} &
\includegraphics[width=0.45\textwidth]{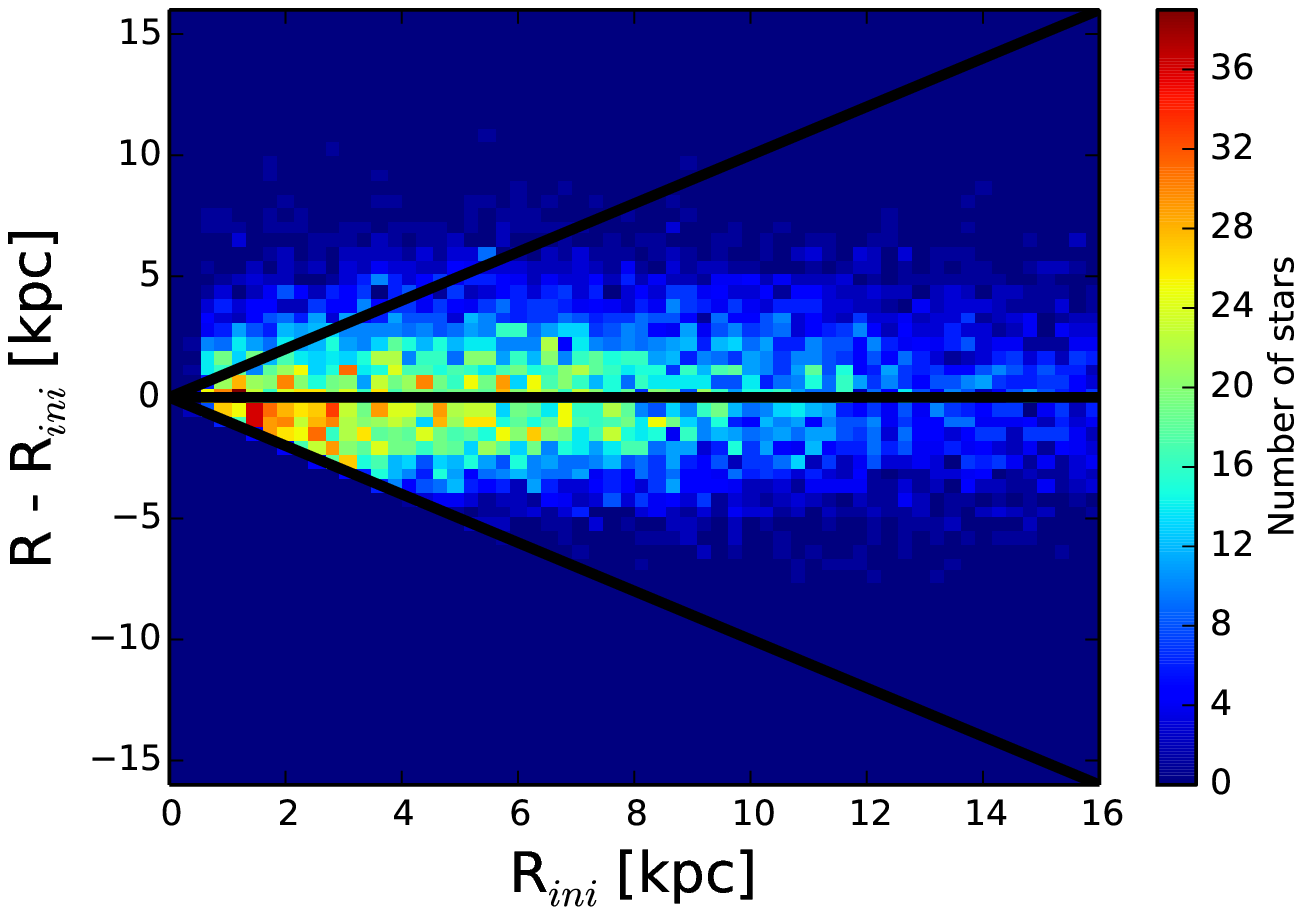} \\
\includegraphics[width=0.45\textwidth]{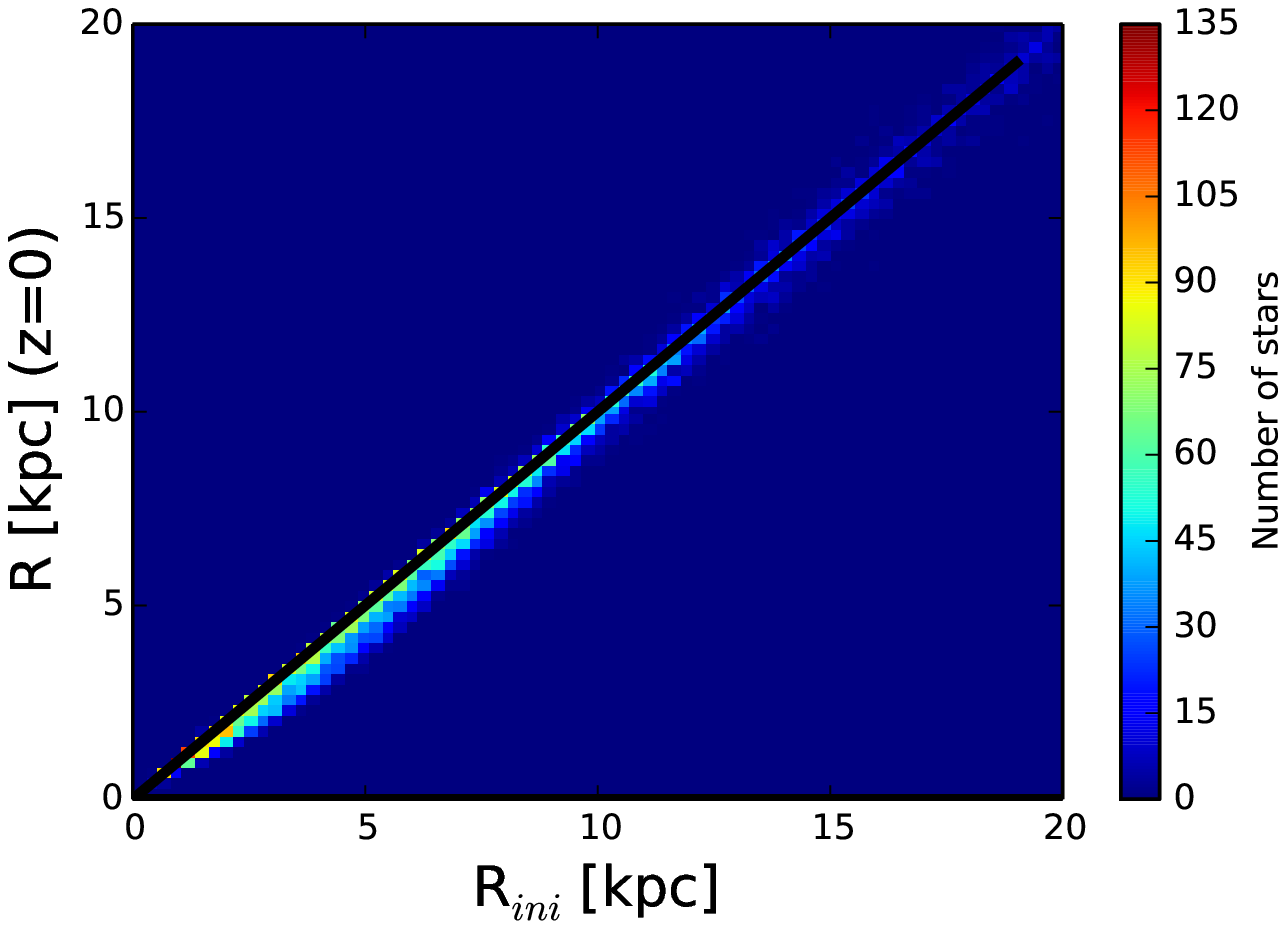} &
\includegraphics[width=0.45\textwidth]{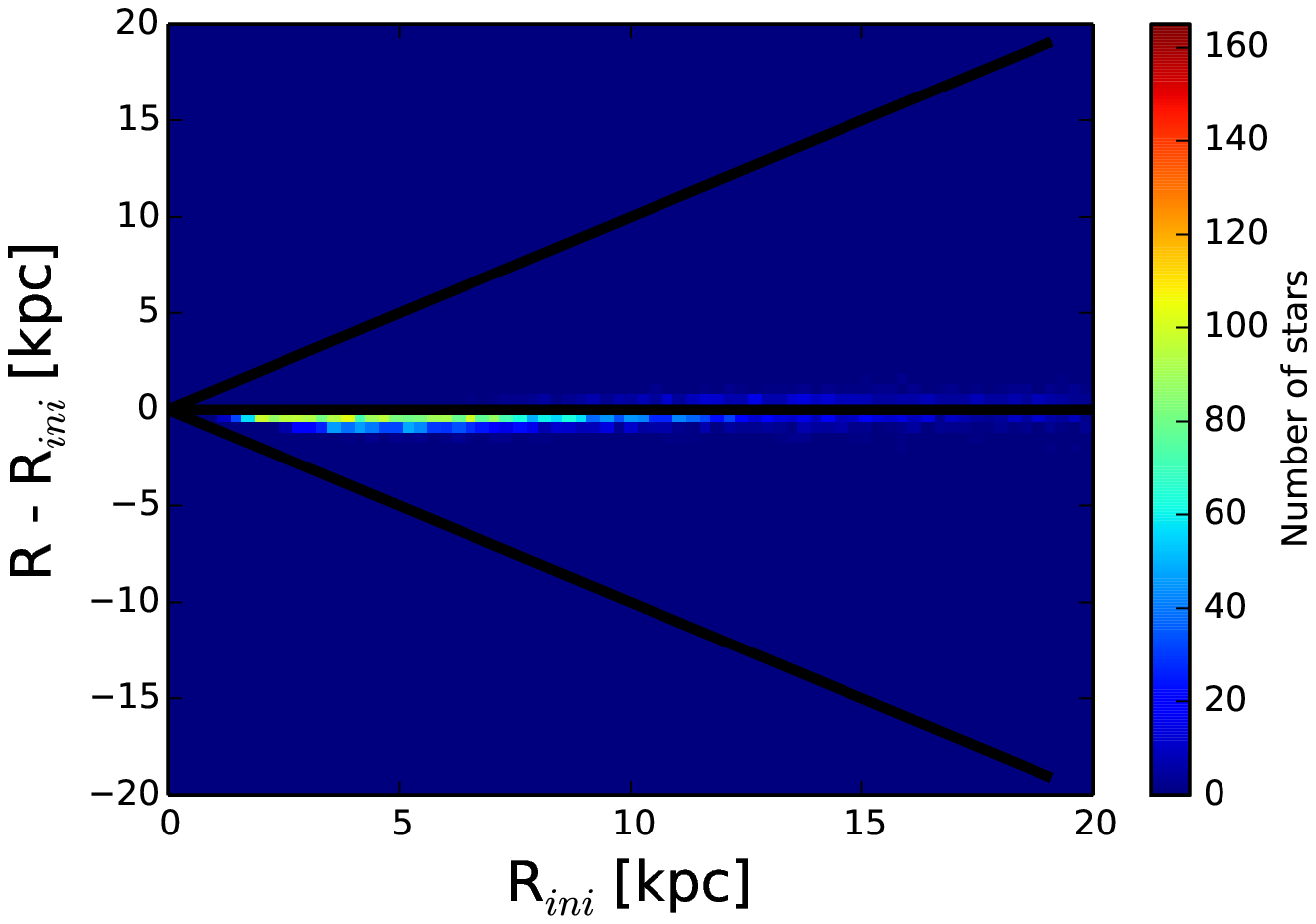}
\end{tabular}
\caption{Initial position versus final positions of the disc stars. Top left panel: initial radii versus final radii, top right panel: initial radii versus the net change in radius, over the 10 Gyr of the simulation. In both top panels diffusion and churning affect the entire disc. Bottom panels: sanity check with churning and diffusion off for initial radii versus final radii (left) and initial radii versus net change in radii (right).}
\label{Migration}
\end{figure*}

In Fig. \ref{Migration}, we show the initial positions of 10\,000 disc stars against their positions after 10 Gyr of integration for two different implementations: {\it{(a)}} churning and diffusion affecting the entire disc and {\it{(b)}} no diffusion or churning. Table \ref{netmigraT} shows the average radial shift that stars have endured under different effects, i.e. potential alone,  churning only, diffusion only and both churning and diffusion. The change in radii ($\Delta R$) for the model with churning and diffusion off, i.e. model `None', merely reflects how the initial disc relaxes owing to the potential only. Even when the heating (represented by the standard deviation given in the table) produced only by churning or diffusion is similar, none of these effects alone is enough to heat the disc up to the level of the observations for all the components, i.e. $U$, $V$, $W$, where $U$ points towards the Galactic Centre, $V$ is in the direction of the  Galactic rotation, and $W$ points towards the Galactic North Pole.

\begin{table}
\caption{Net radial migration (and its standard deviation) of the stars at each of our models, i.e. {\it{None}}: churning and diffusion off, {\it{Churning}}: only on, {\it{Diffusion}}: only on, {\it{Both}}: Churning and diffusion on. Even when the absolute change in radii is similar for all the models, the spread (i.e. migration) of stars represented by the standard deviation is higher when we include both mechanisms: churning and migration.}
\label{netmigraT}
\begin{tabular}{@{}lrc}
  \hline
  Model & $\Delta$R (kpc) & Standard Deviation (kpc) \\ 
  \hline
  None  & $-0.243$ & 0.352\\
  Churning & $-0.194$ & 1.836\\
  Diffusion  & 0.163 & 1.458\\
  Both  & 0.201 & 2.318\\%
  \hline
 \end{tabular}
\end{table}

Comparing the models with and without heating sources, it becomes evident that the mechanisms responsible of this heating, i.e. spiral arms and molecular clouds, drive a significant radial migration of stars.

\section{EVOLUTION OF SMALL OPEN CLUSTERS WITHIN THE GALACTIC DISC}
\label{EVOLUTION OF SMALL OPEN CLUSTERS WITHIN THE GALACTIC DISC}

\subsection{Star clusters models}
\label{cluster-model}

A series of {\it{N}}-body models were run of open clusters by using the code {\tt{NBODY6}} \citep{Aarseth3} with graphical processor unit (GPU) and multicore CPU capability \citep{Nitadori}. For each of our simulations we used six CPU cores (2.66 GHz 64-bit Intel Xeon 5650) and one Tesla C2070 GPU on the GPU Supercomputer for Theoretical Astrophysics Research (gSTAR) at Swinburne University of Technology.

Our primary set of models focused on open clusters consisting of 400 M$_{\odot}$ ($\sim$ 670 stars). Star masses between 0.1 and 50 M$_{\odot}$ were chosen from a Kroupa initial mass function \citep{Kroupa}. We have placed 50 per cent of the stars in our clusters into binary systems, creating a 30 per cent primordial binary frequency, i.e. $\sim$ 330 stars are in binaries in each model. The positions and velocities of the stars within the cluster and the masses of the binary systems were set up as explained in \citet{Guido}. 

The star clusters were evolved by using {\tt{NBODY6}} with a potential for the Milky Way as close as possible to the total ({\tt{GO}}) rotation curve presented in Section \ref{integrator}. For the input of {\tt{NBODY6}} this potential was represented by a time-independent three-component Galaxy, i.e. a point mass bulge with a mass of 6.0 $\times$ 10$^{9}$ M$_{\odot}$, a single \cite{Miyamoto} disc with a mass of 6.0 $\times$ 10$^{10}$ M$_{\odot}$, a scalelength of 2.1 kpc and a scaleheight of 0.3 kpc (which  gives a volume density of 0.05 M$_{\odot}$ pc$^{-3}$ at 8 kpc), and finally a logarithmic halo set by requiring a circular velocity of 220 km s$^{-1}$ at a distance of 8.0 kpc. The rotation curve (solid black curve) in Fig. \ref{velcurve} represent the combined potential used for {\tt{NBODY6}}, i.e. point-mass Bulge, single Miyamoto \& Nagai disc and logarithmic halo. Outside of R$_{\rm{gc}}$ = 3 kpc, where we will evolve our clusters, the potentials from {\tt{NBODY6}} and {\tt{GO}} are indistinguishable in terms of the resultant rotational velocity curves.

\begin{figure}
\includegraphics[width=8.4cm]{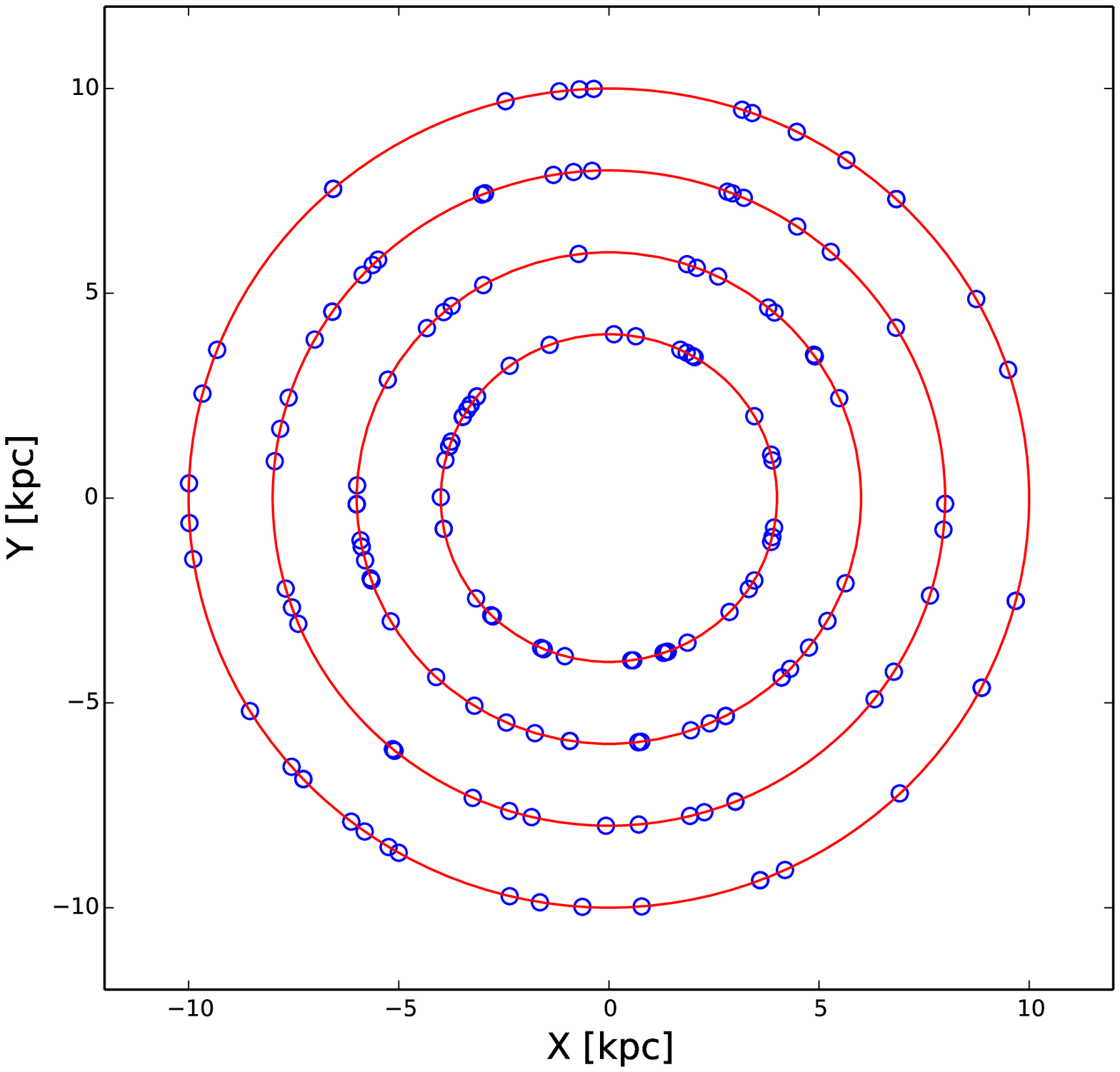}
 \caption{Initial position of our star clusters prior to orbital integration in {\tt{NBODY6}}.}
 \label{IC-clusters}
\end{figure}

We evolved clusters on circular orbits in the disc at 4, 6, 8 and 10 kpc from the Galactic Centre. The initial positions of the clusters were chosen at random azimuthal angles on the orbit, as can be seen in Fig. \ref{IC-clusters}.

There are three main mechanisms that cause stars to escape from stellar clusters: energetic two-body encounters, supernovae events and cumulative two-body encounters (i.e. ``evaporation''). For an analysis of the effect of these three mechanisms on the time-scales and escape velocities of escapers see \citet{Guido}.

At each distance we evolved 40 clusters until at least 70 per cent of the stars in the cluster had escaped. This gives a total of 160 models in our set. We have also evolved a set of 120 models starting with 1\,000 M$_{\odot}$ ($N$ $\sim$ 1\,275) to explore the highly collisional nature of the small evolved clusters and 10 models with 15\,000 M$_{\odot}$ ($N$ $\sim$ 20\,000).

We have integrated the orbits of the stars recorded as escapers from {\tt{NBODY6}} with the orbit integrator previously presented in Section \ref{integrator}. The stars were fed into the {\tt{GO}} integrator as they escaped their parent cluster, which means that stars that escape early during the evolution of the cluster can have up to 400 Myr, i.e. the dissolution time of the cluster, extra-heating with respect to stars that escape at later times.

\subsection{Evolution time-scales}

\begin{table*}
 \caption{Analysis of the systems that escape from our cluster models with 400 M$_{\odot}$ and end up in the Solar annulus 1 kpc wide and centred on 8 kpc. Column 1 represents the radius of the orbit of the cluster before dissolution, {\it{Total $N_{sys}$}} is the contribution of escapers from all the clusters from a given radii, {\it{Systems$_{1 kpc \odot}$}} is the number of these systems that end within the 1 kpc annulus of the Sun after 4.5 Gyr of evolution, {\it{Binaries$_{1 kpc \odot}$}} represents how many of the stellar systems within 1 kpc of the Sun are binaries, and finally column 5 shows this as a percentage. Inside the parenthesis for each column we show the same statistics for ``visible'' stellar systems, i.e. systems that contain at least one main-sequence or giant star.}
 \label{binAnalysis}
 \begin{tabular}{@{}rrrrr}
  \hline
  Radius [kpc] & Total $N_{sys}$ & Systems$_{1 kpc \odot}$ & Binaries$_{1 kpc \odot}$ & Percentage (\%)\\ 
  \hline
   4 & 22200 & 1388\,\,\,\,\,\, (1373) & 451\,\,\,\,\,\, (447) & 32\,\, (32)\\
   6 & 21105 & 3038\,\,\,\,\,\, (2989) & 980\,\,\,\,\,\, (975) & 32\,\, (33)\\
   8 & 19329 & 3679\,\,\,\,\,\, (3615) & 1180\,\,\, (1170) & 32\,\, (32)\\   
  10 & 19109 & 2401\,\,\,\,\,\, (2346) & 746\,\,\,\,\,\, (731) & 31\,\, (31)\\
 {\bf{Total}} &   & {\bf{10506 (10323)}} & {\bf{3357 (3323)}} & {\bf{32 (32)}}\\
  \hline
 \end{tabular}
\end{table*}

\begin{figure}
\includegraphics[width=8.4cm]{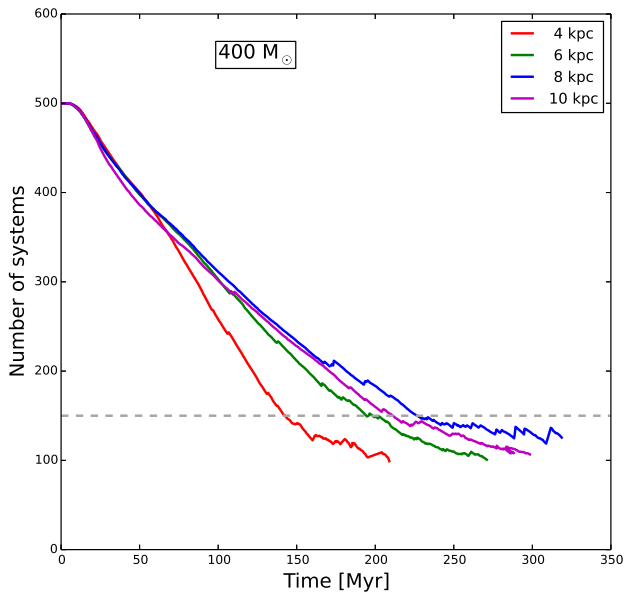}
 \caption{Averaged evolution of the number of stars as a function of time in each cluster at the four Galactocentric distances analysed, i.e. 4, 6, 8 and 10 kpc. The plateau-like regions below the dashed line at 150 systems are an artefact that comes from models that underwent a dissolution greater than 70 per cent  (see Section \ref{cluster-model}).}
 \label{extract-new}
\end{figure}

All our clusters were evolved in the time-independent potential described in Section \ref{cluster-model}. In Fig. \ref{extract-new}, we show how the clusters lose stars over time, where we have averaged over all the models at each Galactocentric distance, i.e. 4, 6, 8 and 10 kpc. As expected, the rate of dissolution of the models at 4 kpc is higher since they feel a stronger gravitational interaction from the central parts of the Galaxy. Models at 6, 8 and 10 kpc `feel' a weaker potential from the inner parts of the Galaxy and all lose members at a lower rate.

The evolution of all these small stellar systems is regulated primarily by their internal evolution, i.e. two-body relaxation time, particularly when the tidal field effect is weak. For our clusters on outer orbits this washes out the marked effect of the Galactic tidal field seen in more massive clusters \citep{Tanikawa2005}.

\begin{figure}
\includegraphics[width=8.4cm]{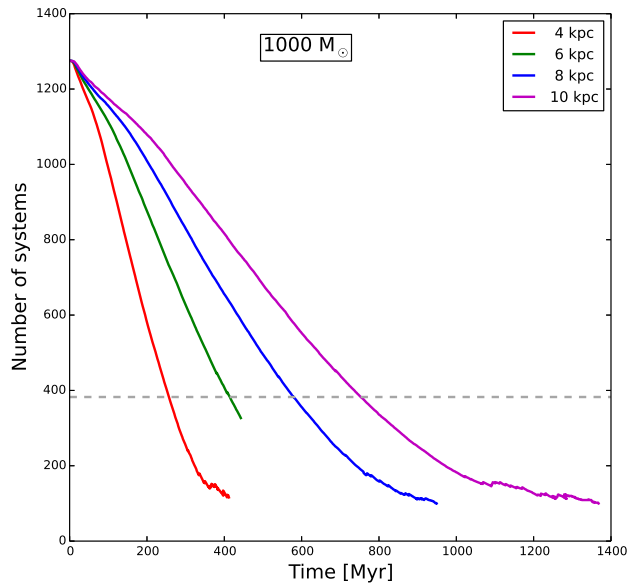}
 \caption{Averaged evolution in time of the number of stars in each cluster of 1\,000 M$_{\odot}$ at the four Galactocentric distances analysed. The horizontal dashed line represents the limit of 70 per cent dissolution (i.e. 380 systems remain).}
 \label{extract-1000}
\end{figure}

A more obvious trend with Galactocentric distance becomes evident when we study more massive clusters, e.g. 1\,000 M$_{\odot}$: see Fig. \ref{extract-1000}. The two-body relaxation time-scale for these more massive clusters is more than twice that of the smaller clusters, which allows them to endure their internal evolution for longer and makes the stripping of stars driven by the Galactic tidal field more evident.

\begin{figure}
\includegraphics[width=8.4cm]{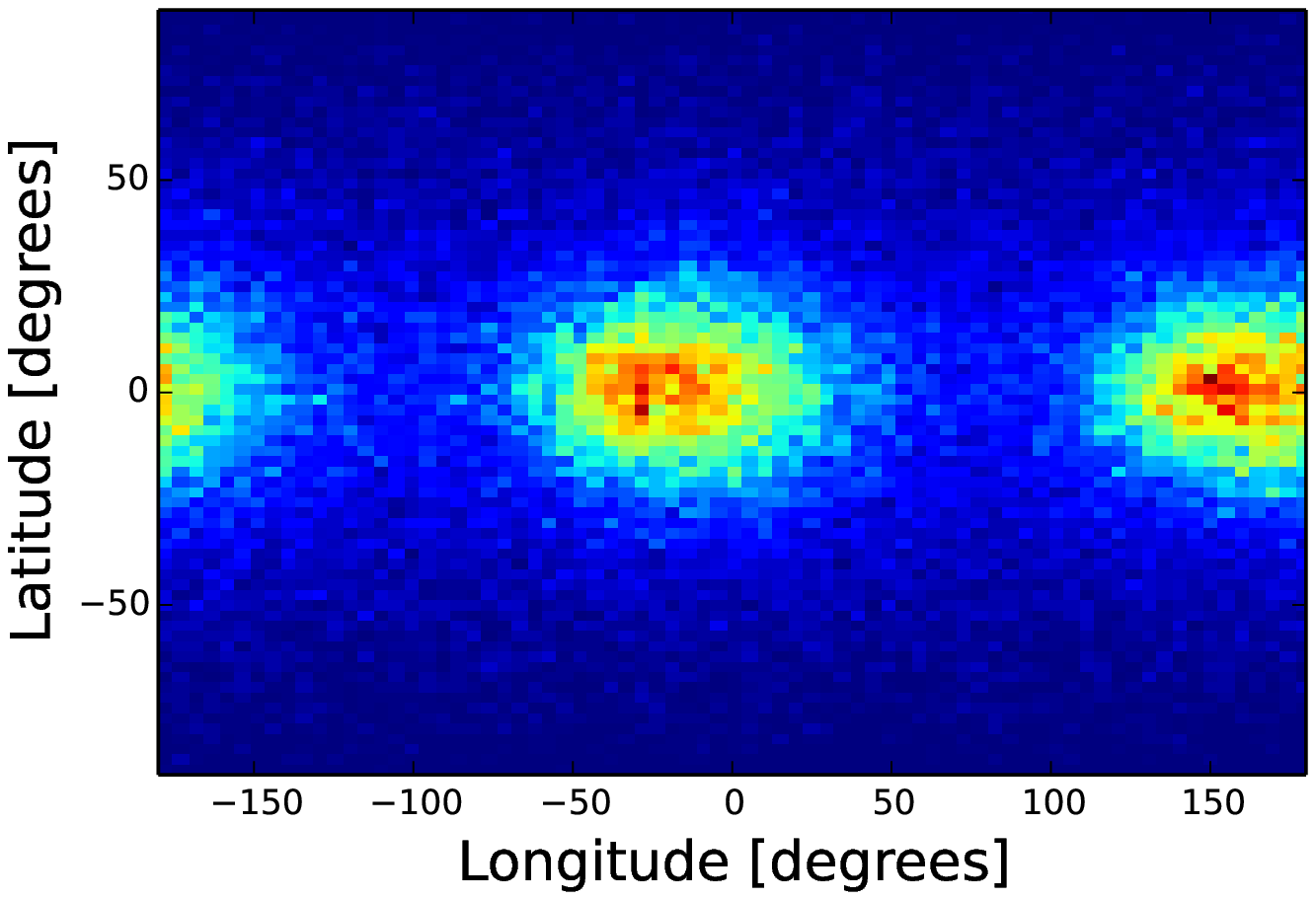}
 \caption{Mercator-like plot showing the directions of the escapers (at twice the tidal radius) after integrating with ({\tt{NBODY6}}). Most of the escapers are produced by evaporation, and the Lagrangian points L1 and L2 are $\sim$ 25 degrees off due to the time where we registered the escapers (twice the tidal radius).}
 \label{Lagrangian-clusters}
\end{figure}

Classical theory \citep{Combes} states that stars in a two-body problem will escape through the Lagrangian points. We have analysed how stellar systems escape from our modelled clusters (see Fig. \ref{Lagrangian-clusters}). From our analysis escaping stars leave the clusters with an offset of $\sim$ 25 degrees from the Lagrangian points L1 and L2. The same result was achieved by \citet{Kupper2008} and \citet{Kupper2011} by analysing the structure of tidal tails of clusters orbiting at different inclinations and eccentricities under the influence of a Milky Way-like potential.

\subsection{Statistics of escaping systems}

Star clusters in the Galaxy lose members through several internal mechanisms which relate to physical processes that range from violent SNe events to dynamical two-body interactions \citep{Guido}. SN ejections are produced within the first 100 Myr of the life of a star cluster. If a given cluster survives the violent switch-off of SNe, two-body interactions become the dominant mechanism that will eject stars.

The number of primordial binaries and particularly the fraction of hard binaries that a given cluster has will directly affect the number of encounters that its members will have before escaping. The higher the primordial binary percentage, the more likely a given star will gain enough energy to escape early in the evolution of its parent cluster (\citealt{Hut}; \citealt{Guido} and references therein). 

In the context of chemical tagging and reconstruction of dissolved populations (\citealt{DeSilva2009}; \citealt{Joss2010}; \citealt{DeSilva2013}) binaries play an important role given that the members of these systems are more likely to have their surface abundances altered by mass transfer in the case of highly eccentric orbits or during Roche-Lobe overflow, provided the binary system lives long enough to allow one of its components to evolve as a giant (\citealt{DeDonder2002}; \citealt{Vanbeveren2007}). Furthermore, since binary systems fill up to 50 per cent of the systems in the solar neighbourhood recent studies have brought to attention the importance of studying the habitability zones of these systems (\citealt{Jaime};  \citealt{Eggl}).

Escaping binary systems represent 30 per cent of the total number of systems that our disrupted 400 M$_{\odot}$ clusters contribute to the Galaxy. This value is maintain from the 30 per cent primordial binary fraction that we have assumed and we note that within these escaping binary systems we are including both a fraction of  primordial binaries that survived the dynamical evolution of the cluster and binary systems that were created by multiple-body encounters.

We have analysed the number of binary systems that reached the solar annulus after 4.5 Gyr with an origin from clusters at different Galactocentric distances. As can be seen in Table \ref{binAnalysis}, of all the stellar systems (single stars plus binary systems) that escaped from our clusters and ended within the Solar annulus, roughly 30 per cent of them are binaries. This means that of the systems escaping from the clusters, there is no clear preference for either single stars of binaries reaching the Solar neighbourhood.

We can also infer from Table \ref{binAnalysis} the contribution of possible solar siblings, i.e. stars born in the same stellar nursery as the Sun, depending on our explored birth locations, e.g. 4, 6, 8, 10 and 12 kpc. From clusters born at 8 kpc the contribution is 17 per cent higher than from clusters at 6 kpc and 35 per cent higher than clusters at 10 kpc. Binary systems in the solar annulus have similar fractional contributions from each family of clusters orbits, but the larger contributors in numbers of binaries within the solar annulus are the clusters born within it. Binary systems from clusters born at 8 kpc outnumber binary systems born at 6 kpc by 17 per cent and binary systems born at 10 kpc by 37 per cent. We expect this to be the case since the escape velocity of binaries is lower than single stars owing to their main escaping mechanism being evaporation \citep{Guido}, making them stay longer near the orbit of their parent cluster.

\citet{Duquennoy} found that the frequency of binary systems in the solar neighbourhood is close to 50 per cent, whereas the contribution from our clusters is $\sim$ 30 per cent (as noted above) which is low if we were to assume that all stars originated in clusters. 
For our primordial binaries, we assign a period ($P$) cut-off of $P$ < 10$^{7}$ d. 
If this cut-off were to be relaxed to longer periods, e.g. $P$ $\sim$ 10$^{10}$ d, we could effectively have a greater primordial binary fraction by starting more single stars as binaries. 
However, a higher number of wide primordial binary systems will enhance the number of interactions owing to their larger gravitational cross-section which will enhance the disruption of these additional soft binaries. 
Conversely, we could simply increase our primordial binary percentage to 50 per cent using the same period distribution as before and it would be reasonable to expect that our escaping binary frequency would then be a similar value. 
We also need to consider that even with 30 per cent of escapers being binaries we might already be overestimating the fraction of binaries within the solar annulus because our value is calculated by including all the binaries regardless of their mass ratio or brightness. 
Plus any wide systems that escape from the parent clusters could be hard to be identified as binaries in the field owing to their separations. 
A further analysis \--- outside the scope of the present work \--- which includes different period cuts limited to observational capabilities such as SDSS and LAMOST \citep{Shuang} and mass ratio (\citealt{Duquennoy}; \citealt{Parker}) is required.

\subsection{Losing information: birth places of dissolved clusters}
\label{Losing information: birth places of dissolved clusters}

\begin{figure*}
\centering
\begin{tabular}{cc}
 \includegraphics[width=0.45\textwidth]{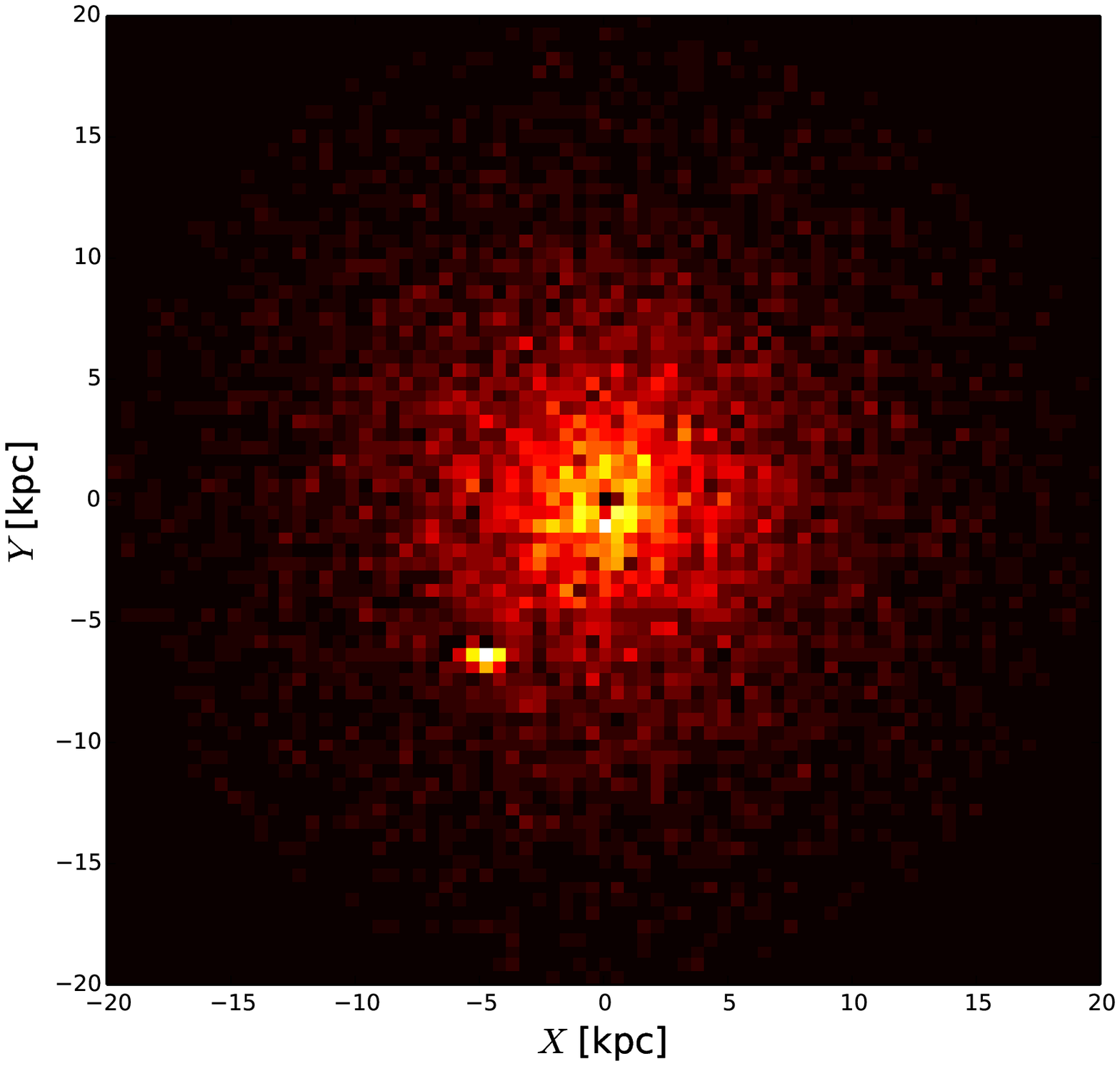} &
 \includegraphics[width=0.45\textwidth]{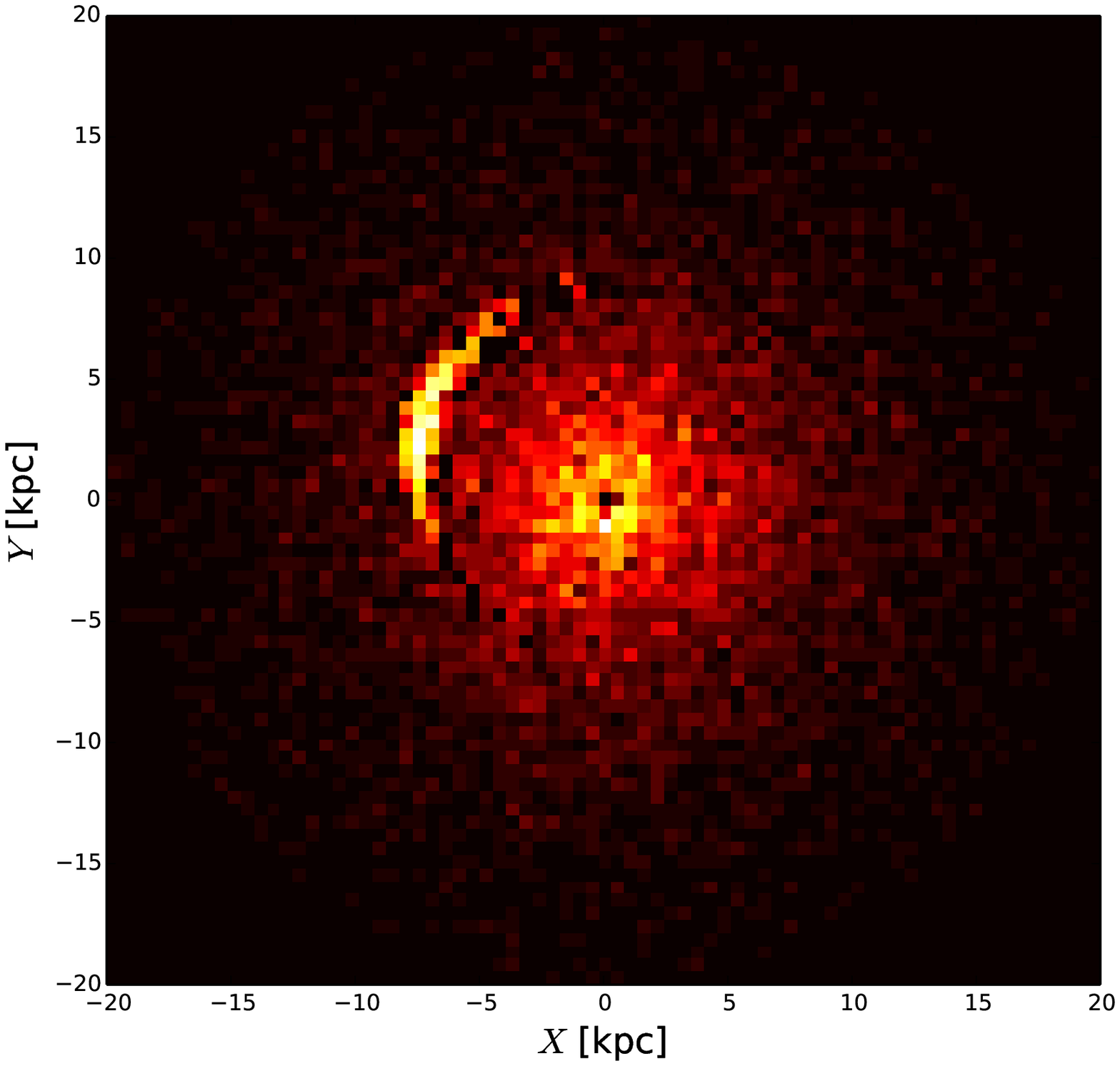} \\
 \includegraphics[width=0.45\textwidth]{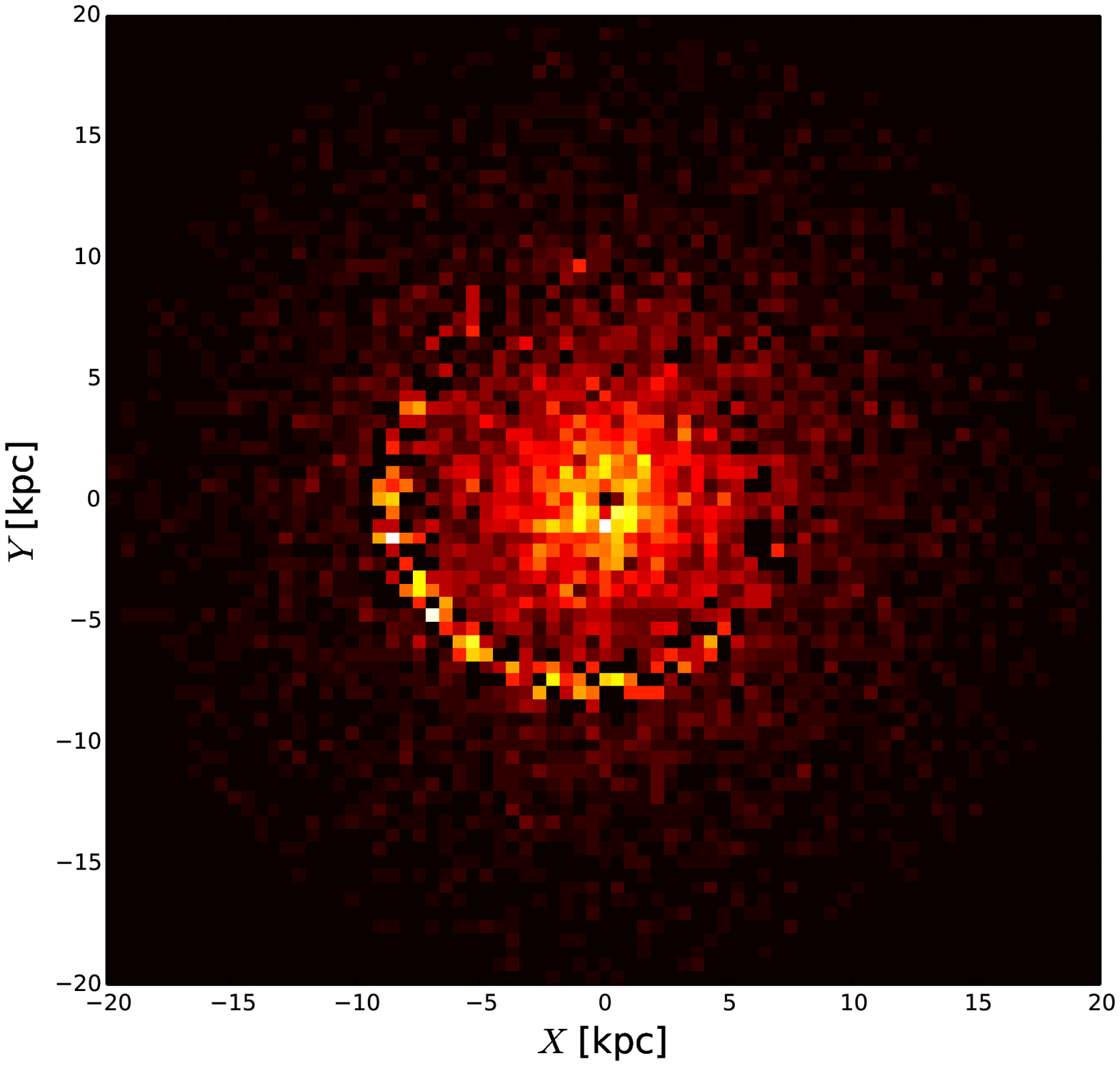} &
 \includegraphics[width=0.45\textwidth]{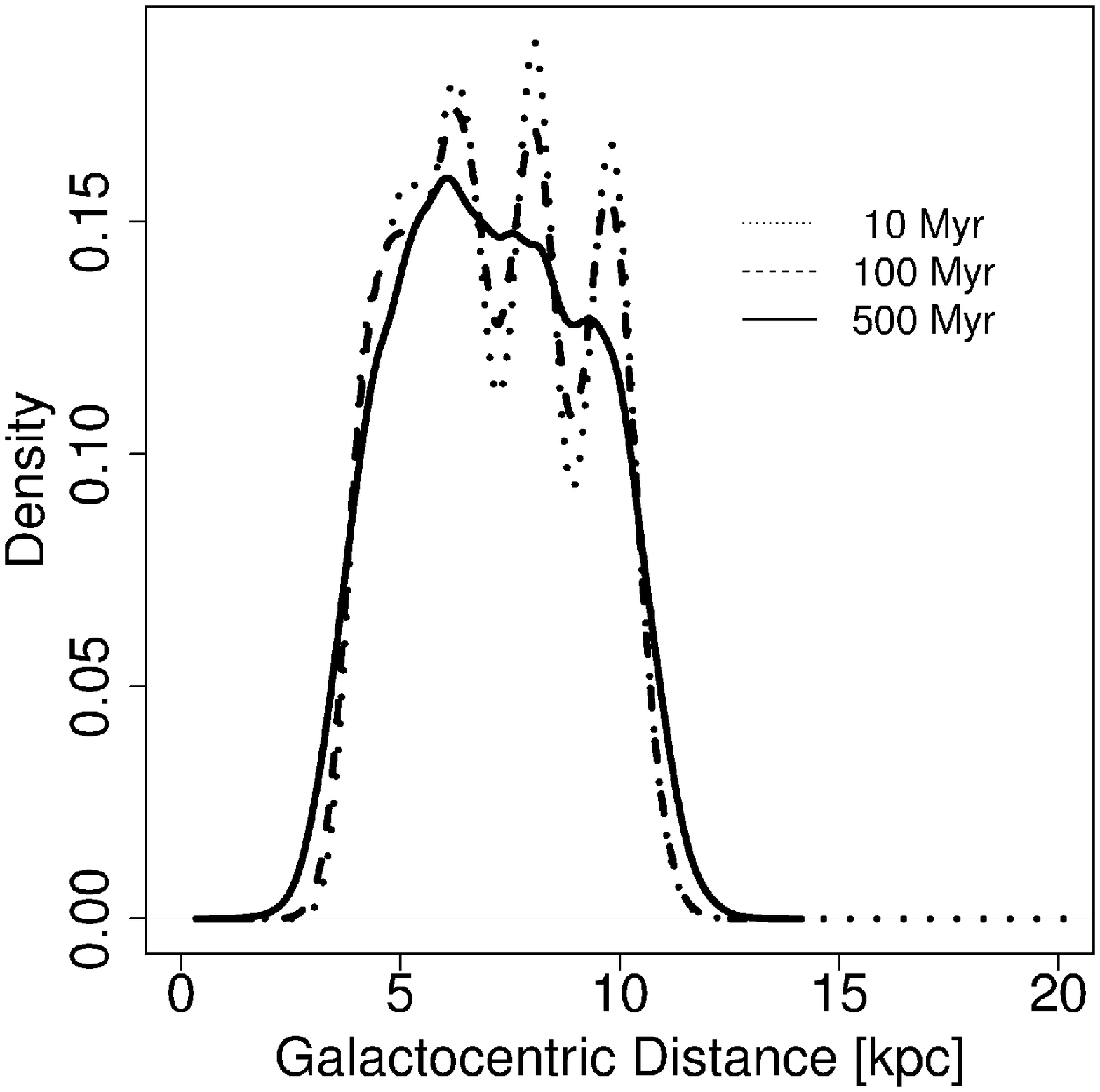}\\
\end{tabular}
\caption{Configuration space for our Galactic disc and one particular cluster starting with 400 M$_{\odot}$ at different degrees of dissolution, i.e. top-left: 100 Myr of evolution, top-right: 500 Myr of evolution, bottom-left: 1 Gyr of evolution. Bottom-right: evolution of the radial density distribution for all our clusters that started with 400 M$_{\odot}$ at three times: 10 Myr, 100 Myr and 500 Myr after dissolution. Evident is the smoothing of the overdensities, i.e. the decrease of the density peaks, as time passes.}
\label{finder-configuration}
\end{figure*}

By following stars as they drift away from the cluster and join stars in the field we can aim to understand  the possibility of reconstructing these dissolved clusters, i.e. can we still distinguish the clusters members for some length of time past dissolution?

We begin by studying the distribution of escapers from a single cluster born at 8 kpc at different stages of dissolution, e.g. after 100 Myr of evolution, after 500 Myr of evolution and after 1 Gyr of evolution. This particular cluster dissolves completely after 370 Myr.

The natural starting place to try to find this cluster is real space ($X$-$Y$ space), hereafter `configuration space'. We have enhanced the density contrast of colours in Fig. \ref{finder-configuration} to facilitate the identification of the cluster but it is clear from the figure that the cluster will be well mixed with the field stars in $\sim$ 500 Myr. Clustering footprints could last longer in phase-space especially for massive satellites in the halo (\citealt{Gomez2010a}; \citealt{Gomez2010b}). Since the density contrast of stars in the Galactic disc is higher than the density in the halo, most of the techniques used for finding phase-space structures on the Galactic halo are not applicable to the Galactic disc. We have performed an initial investigation on the signatures of disrupted clusters in phase-space. From our findings it seems possible to identify disrupted clusters with initial 400 M$_{\odot}$ up to 500 Myr after dissolution. Regardless of how the initial clusters are distributed (here, essentially an anti/inverted-exponential for the 6 to 8 kpc clusters) after $\sim$ 500 Myr, the dissolved cluster debris relaxes into an exponential surface density distribution consistent with that of the background field stars. In the future we will perform a full phase-space analysis (Moyano Loyola et al., in prep.).

In Fig. \ref{finder-configuration} we show the radial density distribution of our dissolved 400 M$_{\odot}$ clusters at three different times: 10 Myr, 100 Myr and 500 Myr after dissolution. When we refer here to a given time after dissolution, we have considered that each cluster dissolves at a slightly different rate, so we have calculated the mean dissolution time for each family of  clusters at each Galactocentric distance. In this figure it is evident how the overdensities corresponding to the four birth radial locations of this set, i.e. 4, 6, 8 and 10 kpc, merge into a single smooth distribution in only 500 Myr. After this stage chemical information becomes indispensable for reconstructing star clusters \citep{Joss2010}. Future papers in the Tracking Cluster Debris (TraCD) series will include a multi-dimensional group finding approach within chemistry-space, as part of our test of chemical tagging (Macfarlane, Gibson, et al., in prep). It will also be of interest to further investigate the dissolution rates and over-density behaviour in light of comparable previous work (e.g. \citealt{Chumak2005}) in our next stages.

The time-scale in which the overdensities disappear is closely related to the dissolution time-scale, i.e. the time-scale for the smoothing of overdensities for our models with 1\,000 M$_{\odot}$ is more than twice the value for our 400 M$_{\odot}$ models. This scaling is also representative of the relative half-mass relaxation time-scales of the clusters which is again roughly a factor of two longer in the more massive clusters.

\section{THE SEARCH FOR THE SUN}
\label{The search for the Sun}

There is increasing evidence that stars are born generally in groups (\citealt{Tutukov};  \citealt{Carpenter}; \citealt{Lada}; \citealt{Porras}; \citealt{Bressert}; \citealt{Kruijssen}). Signatures such as the excitation of some minor bodies in the Kuiper belt, the presence of short-lived radioactive isotopes in meteorites and the orbits of external giant planets like Uranus and Neptune suggest that the Sun is no exception (\citealt{Gaidos1995}; \citealt{Gaidos2009}; \citealt{Adams}; \citealt{Jimenez-Torres}; \citealt{Pichardo}; \citealt{Pfalzner}).

It has been suggested that the nearby Galactic open cluster M67 is a candidate for the birth place of the Sun due to the similarities between the chemical composition of stars in this cluster and the Sun (\citealt{Friel}; \citealt{Tuat}; \citealt{Randich}; \citealt{Pace}; \citealt{Pasquini}; \citealt{Onehag}; \citealt{Castro}). However, recent studies have put this idea into doubt \citep{Pichardo}. The identification of the parent star cluster where the Sun was born and particularly the possibility of finding solar siblings, i.e. stars that were born in the same star forming region as the Sun, is of crucial importance to unravelling the dynamical history of the Galactic disc during the last $\sim$ 5 Gyr. This information will improve constraints on the birth environment of the Sun, i.e. number of members, mass, size, as well as the potential of the Milky Way.

It would be natural to assume that since we have a good estimate of the velocity of the Sun a sensible starting point might be to just invert its velocity vector and use any orbit integrator to integrate `backwards' in time for $\sim$ 4.5 Gyr (i.e. the age of the Sun) to find its birth place within the disc. \citet{SimonSiblings} conducted such an analysis by utilizing the static, i.e. no time-dependent features, three-component potential described by \citet{Paczynski}. The main results showed how the probability of finding siblings amongst unrelated field stars diminishes as time passes along the orbital trajectory of the Solar system in the Galaxy. As we have discussed in Section \ref{STRUCTURE OF THE DISC}, the presence of spiral arms and GMCs imposes a non-reversible time feature to the potential of the Milky Way making the backwards integration not a suitable approach to study orbits within the Galactic disc.

\begin{table*}
 \caption{Main sequence stars with mass in the range 0.8$-$1.2 M$_{\odot}$ that ended within the Solar annulus after 4.5 Gyr from all the modelled clusters. {\it{Radius}} is the birth orbit of the parent cluster, {\it{Stars$_{1 kpc \odot}$}} is the number of stars that ended within the solar annuli, {\it{Circular orbit}} is the number of stars that ended within the solar annuli and have circular velocities between 200-240 km s$^{-1}$ and {\it{Radial velocities}} shows how many of those stars have radial velocities lower than 15 km s$^{-1}$.}
 \label{raiders}
 \begin{tabular}{@{}cccc}
  \hline
  Radius [kpc] & Stars$_{1 kpc \odot}$ & Circular orbit & Radial velocities $\le$ 15 km s$^{-1}$\\ 
  \hline
  & & 0.8 M$_{\odot}$ $\le$ mass $\le$ 1.2 M$_{\odot}$ & \\
  \hline
   &     &   400 M$_{\odot}$ clusters\\
  \hline 
   4 & 46 & 13 & 2 \\
   6 & 142 & 82 & 23\\
   8 & 150 & 96 & 45\\   
  10 & 81 & 62 & 28\\
  \hline
   &     &   1\,000 M$_{\odot}$ clusters\\
  \hline
   4 & 105 & 18 & 2\\
   6 & 229 & 121 & 32\\
   8 & 255 & 161 & 64\\   
  10 & 168 & 119 & 48\\
  \hline
   &     &   15\,000 M$_{\odot}$ clusters\\
  \hline
   4 & 78 & 0 & 0\\
   6 & 164 & 11 & 5\\
   8 & 284 & 30 & 17\\   
  10 & 68 & 12 & 5\\
  12 & 75 & 15 & 7\\
  \hline
 \end{tabular}
\end{table*}

The orbit of the external giant planets, i.e. Uranus and Neptune, the orbit of the large planetoid Sedna and the eccentricities, inclinations and truncation at $\sim$ 50 au of the orbits of the objects within the Kuiper belt place constraints on the central star density of the star cluster containing the Solar system to be in the range 10$^{3}$ $\leq$  $\rho_{c}$ $\leq$ 10$^{5}$ M$_{\odot}$pc$^{-3}$ (\citealt{Gaidos1995}; \citealt{Adams}; \citealt{Pfalzner}). The range corresponds to limits on the number and strength of close encounters (or the lack of them) that could have disrupted the protoplanetary disc of the Solar system (\citealt{Adams}; \citealt{Jimenez-Torres}). The position of the Sun in such a cluster is also a factor, given that it would need to reside in the outer regions of the cluster at the higher end of the density range in order for the Solar system to survive intact. These considerations place an upper limit of $N$ $\sim$ 10$^{5}$ stars for the birth cluster.

Another constraint comes from the presence of short-lived radionuclides (SLRs: $^{10}$Be, $^{26}$Al, $^{36}$Cl, $^{41}$Ca, $^{53}$Mn, $^{60}$Fe, $^{107}$Pd, and $^{182}$Hf) in the early Solar system found nowadays in meteoritic materials (\citealt{Lugaro} and references therein). Two origins have been proposed for these SLRs: Type II SN (which will lead to a chemically inhomogeneous cluster) and Wolf-Rayet stars (chemical homogeneous). The SN scenario requires a progenitor with M$_{\ast}$ $\approx$ 25 M$_{\odot}$ or more, and thus requires a cluster with at least $N$ $\approx$ 825 stars to have a 50/50 chance or better of hosting a 25 M$_{\odot}$ star (\citealt{Adams}; \citealt{Pfalzner}). The Wolf-Rayet scenario requires an even more massive star with M $\approx$ 60 M$_{\odot}$ which translates to a more massive cluster, i.e. a cluster with N $\ge$ 10$^{4}$ stars (\citealt{Gaidos2009}; \citealt{Adams}).

Finally, the UV radiation field of the background of the cluster often dominates the radiation provided by the star itself. This UV field can lead to photoevaporation of protoplanetary discs. The solar nebula is then constrained to be weak enough to allow gas to survive up to radii $\leq$ 30 au, i.e. the orbit of Neptune \citep{Adams}.

The combination of these constraints from these studies suggests that our model clusters starting with 400 M$_{\odot}$ are very much at the low end of the spectrum for the mass (or $N$) of the birth cluster. We will also utilize our models starting with 15\,000 M$_{\odot}$ ($N$ $\sim$ 20\,000) to explore clusters at the upper end of the spectrum and our models starting with 1\,000 M$_{\odot}$ which sit in the most likely region of the spectrum. As well as looking for Solar-candidates the data we present are also applicable to solar twins, i.e. stars that are particularly similar to the Sun in effective temperature, metallicity and age. The solar neighbourhood can easily be populated by a non-negligible fraction of solar twins.

Using our time-dependent potential (see Section \ref{disc-model}), we have analysed the possibility that an  escaping star from our modelled clusters with a mass similar to the Sun ended on a circular orbit within a 1 kpc annulus centred at 8 kpc after 4.5 Gyr, becoming a Sun-candidate. Our tests involved all clusters of 400 M$_{\odot}$, 1\,000 M$_{\odot}$ and 15\,000 M$_{\odot}$. In Table \ref{raiders} we show the number of stars from the clusters born at each Galactocentric radii that ended between 7.5 $\leq$ R$_{\rm{gc}}$ $\leq$ 8.5 kpc, from those stars how many have a circular velocity within 10 per cent of $V_{\odot}$ = 220 km s$^{-1}$, and how many of those stars in a  circular orbit have a radial velocity $U \leq 15$ km s$^{-1}$.

In terms of probabilities, we have found that for our 400 M$_{\odot}$ clusters there is a 49 per cent higher chance of the Sun being born from a cluster at 8 kpc than at 6 kpc while only a 38 per cent higher chance relative to clusters born at 10 kpc. For our 1\,000 M$_{\odot}$ clusters there is a 50 per cent higher chance of the Sun being born from a cluster at 8 kpc than at 6 kpc while only 25 per cent higher chance relative to clusters born at 10 kpc. Finally for our 15\,000 M$_{\odot}$ clusters there is a 70 per cent higher chance of the Sun to be born from a cluster at 8 kpc than at either 6 or 10 kpc while only a 59 per cent higher chance relative to clusters born at 12 kpc.

From Table \ref{raiders}, we can see that even for the low-mass clusters in our analysis, i.e. 400 and 1\,000 M$_{\odot}$, it is possible for a star similar to the Sun to end on a Sun-like orbit after 4.5 Gyr. The most likely Sun-candidates will be stars coming from clusters that were born at 8 kpc which is in agreement with the dynamically inferred results for the solar birth location in \citet{Minchev} and \citet{Martinez-Barbosa}.  Stars from clusters born at 4 kpc can reach the Solar orbit in 4.5 Gyr but this is much less likely.

The number of Sun-like stars reaching the solar orbit from the  15\,000 M$_{\odot}$ clusters is lower owing to the fact that these clusters are more resilient to stripping of their members through the tidal interaction with the Galaxy. The longer lifetime of these clusters corresponds to a lower mass-loss rate which in turn translates into a lower proportion of stars that could reach the solar annulus within 4.5 Gyr. For these more massive clusters, the larger contribution comes once again from the clusters born at 8 kpc.

As well as looking for 
solar-candidates from our models, i.e. stars that have escaped from a star cluster and ended at 8 kpc with circular velocity equal to 220 km s$^{-1}$ and within the mass range 0.8$-$1.2 M$_{\odot}$, 
Table \ref{raiders} also gives an indication of the numbers of potential solar-twins that may populate the Solar neighbourhood. 
When observing such stars, if we are not careful we might confuse solar-siblings (from the same birth cluster as the Sun) with Solar-twins (appear similar but from a different birth cluster). The question then arises: {\it{how can we break the degeneracy between Solar-twins and Solar-siblings stars in such cases?}} The answer most likely comes from analysing chemical space. 

A key required property of the Solar Nebula is its chemical homogeneity. If the Solar-siblings share the same chemical composition it will be possible to find them in chemical surveys like GALAH-AAO\footnote{http://www.mso.anu.edu.au/galah/home.html}. Clusters up to 10$^{4}$ M$_{\odot}$ and a significant fraction of clusters up to 10$^{5}$ M$_{\odot}$ are expected to be chemically homogeneous \citep{Joss2010}. 

The previously mentioned homogeneity of the Solar Nebula is closely related to the enrichment mechanism of the cluster. If a SN trigged the star formation that created the Sun, then significant chemical inhomogeneity was injected into the stars formed from that solar nebula. If the contaminant was a Wolf-Rayet star, the contamination would occur prior to the formation of most of the solar-siblings (and the Sun itself) which will make the cluster chemically homogeneous. Lastly, chemical homogeneity can be reached if the SLRs were accreted after star formation in the cluster, but before cluster dissolution \citep{Joss2010}.

\subsection{Observational samples: {\it{Gaia}} and GALAH}

Three large surveys will provide unprecedented information about the assembly history and current state of the Milky Way and the location of the solar family: {\it{Gaia}} \citep{Perryman}, {\it{Gaia}}-ESO \citep{Gilmore} and GALAH-AAO (Galactic Archaeology with HERMES-Australian Astronomical Observatory; see \citet{ZuckerGALAH} for a review on the project). {\it{Gaia}} will provide high-precision distances and proper motions and will synergies with {\it{Gaia}}-ESO which will use FLAMES@VLT to obtain high precision spectroscopic data. Finally GALAH-AAO is focused on finding dissolved star clusters and associations by tagging chemical elements. 

The {\it{Gaia}} surveys will be magnitude limited, with $G \leq$ 20 for {\it{Gaia}}, $V \le 19$ for faint stars and $ V \le 16.5$ for bright stars for {\it{Gaia}}-ESO. GALAH-AAO in addition to a magnitude cut of 12 $\leq V_{\rm{2MASS}} \leq$ 14 will observe stars with $E(B-V)$ $<$ 0.2 and Galactic latitudes more than 5 degrees ($|b|>5^{\circ}$) off the plane.  The Johnson colours can be converted to $G$-band and $V_{\rm{2MASS}}$ following \citet{Jordi} and \citet{Bilir} respectively. For our analysis we work with $V$ from the Johnson system \citep{Johnson} and assume $V_{\rm{2MASS}} \sim V$.

We mainly aim to model the dynamics of dissolved star clusters and the field disc stars where all these clusters will mix, so now we need to translate our dynamical variables to observations in order to be able to make predictions about these upcoming surveys.

The first selection implemented was to only consider stars with masses up to 1.33 M$_{\odot}$.  Stars up to this mass limit will not evolve beyond the asymptotic giant branch in 4.5 Gyr of evolution. This selection was needed because we wanted to consider only stars that will remain visible at the moment of taking our mock samples at 4.5 Gyr.

In order to calculate the absolute magnitudes of our samples at 4.5 Gyr we first take the mass of the stars when they escape from the {\tt{NBODY6}} model and evolve each escaping star with the Single Stellar Evolution ({\tt{SSE}}) code \citep{Hurley1} to retrieve their stellar properties, i.e. current mass, radius, evolutionary stage and luminosity, at an age of 4.5 Gyr.
We then use these properties to calculate the absolute magnitudes in the Johnson system \citep{Johnson}  by using stellar atmosphere models \citep{kurucz} and the appropriate distance modulus (see below) to calculate the apparent magnitude of our modelled stars. Finally, we included the effects of dust in the Galactic disc by using the model presented in \citet{Hakkila}.

To improve our statistics, we took advantage of the fact that our Galactic model is axisymmetric, so we have chosen four zones in our Galactic disc from which we selected our samples: (8,0,0); (0,8,0); ($-8$,0,0) and (0,$-8$,0). We have centred the Sun on each of these points and calculated the corresponding distance modulus and extinction to each of our stars, limiting the sample to $V$ $\leq$ 20, i.e. to the faintest magnitude limit of both of the {\it{Gaia}} surveys mentioned before.

\begin{figure}
\includegraphics[width=0.45\textwidth]{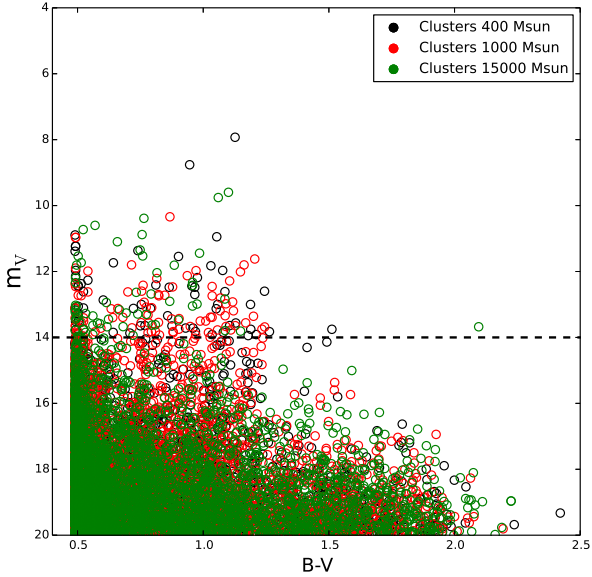}
\caption{Colour-Magnitude diagram using the Johnson visual magnitude ($V$) after 4.5 Gyr. The sample of stars was limited to $V \leq$ 20 around the Sun. The symbols are colour coded according the mass of the parent cluster. The dashed line indicates the $V$ limit for the GALAH survey, i.e. $V \leq 14$.}
\label{magnitudes1}
\end{figure}

\begin{figure}
\includegraphics[width=0.45\textwidth]{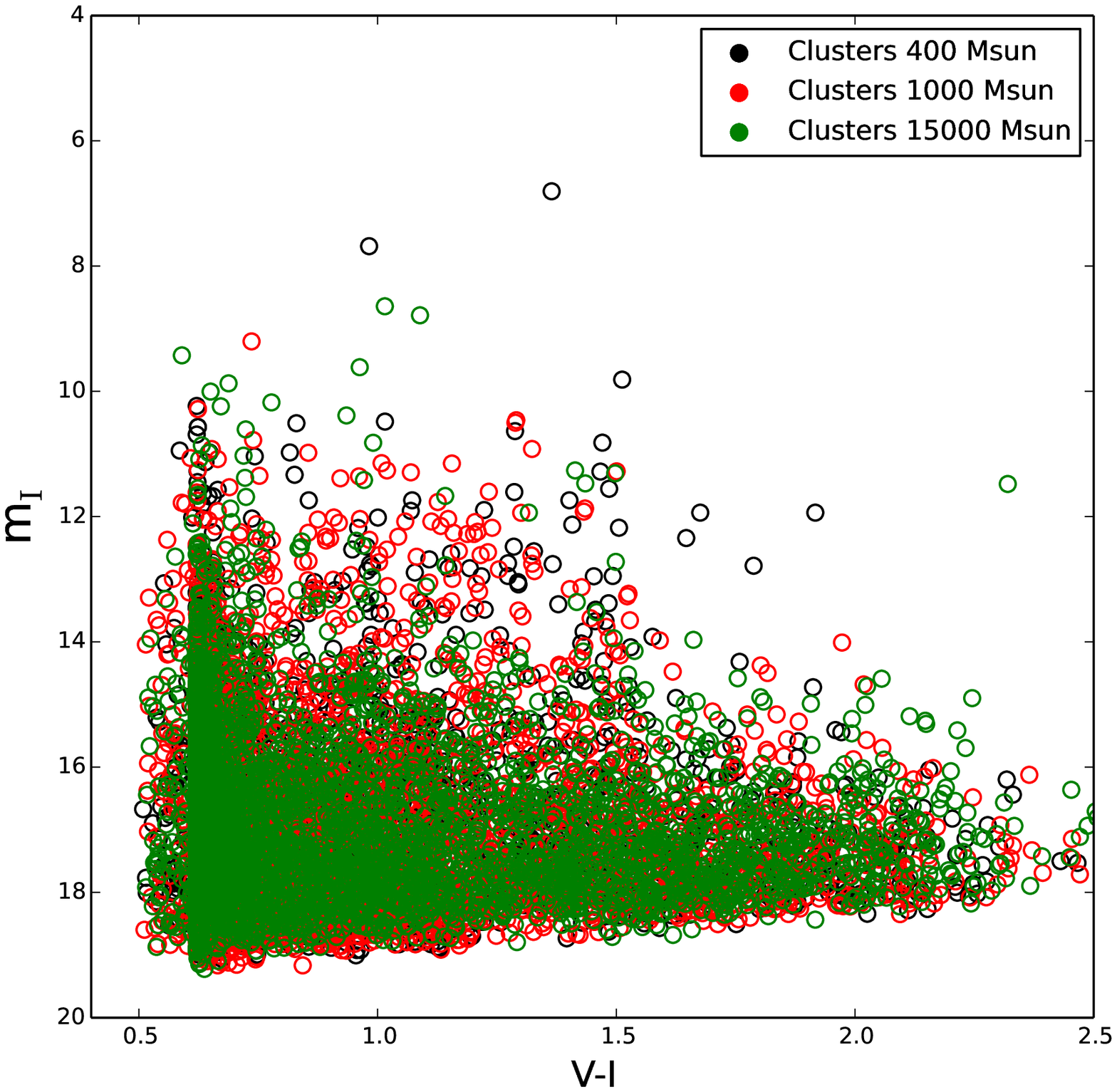} \\
\caption{As Fig. \ref{magnitudes1} but for the $I$-band.}
\label{magnitudes2}
\end{figure}

\begin{figure}
\includegraphics[width=0.45\textwidth]{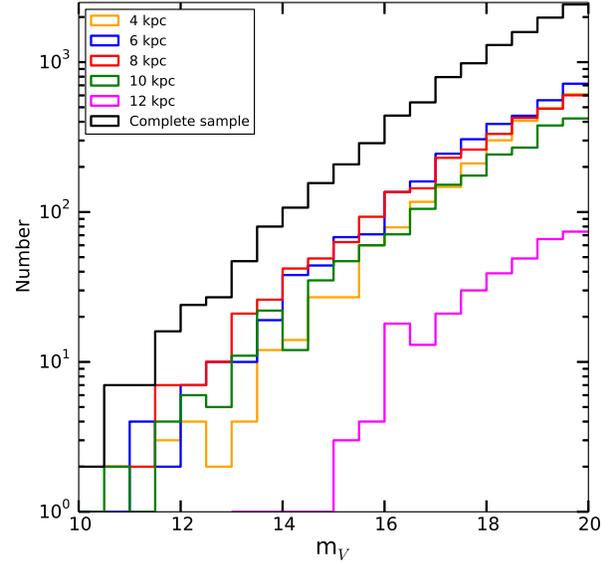} \\
\caption{Histograms of the distribution of $V$ magnitudes of our sampled stars from the distribution of cluster masses in Fig. \ref{magnitudes1}. Here we have separated the sample by the birth location of the cluster.}
\label{mvhist}
\end{figure}

\begin{figure}
\includegraphics[width=0.45\textwidth]{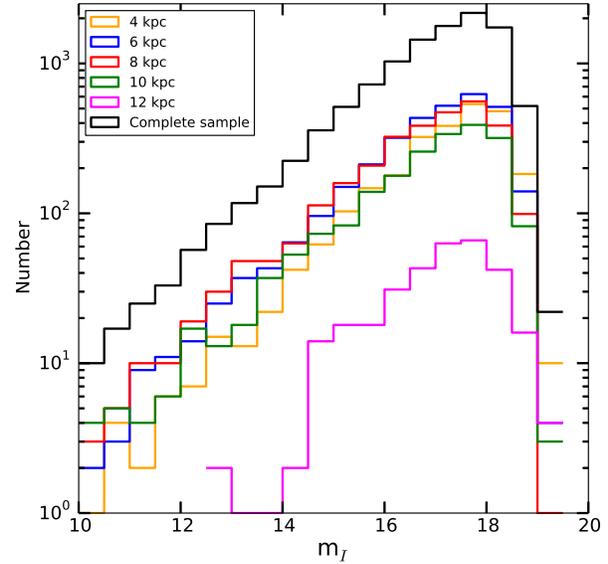} \\
\caption{Same as Fig. \ref{mvhist} but for the $I$-band.}
\label{mihist}
\end{figure}

In Figs. \ref{magnitudes1}--\ref{magnitudes2} we present colour-magnitude diagrams (CMDs) for this sample of stars that have escaped from dissolved star clusters. In Fig. \ref{magnitudes1} we also indicate the limit where $V \leq $ 14, i.e. roughly the limit of GALAH-AAO. In Figs. \ref{mvhist}--\ref{mihist} we show the histograms of the distributions of $V$ and $I$ colours for our samples. In these histograms we can see that stars escaping from clusters born at 6 and 8 kpc will contribute in a similar way to these observational samples, followed by contributions from clusters born at 4 kpc and finally a smaller contribution from clusters born at 12 kpc.

In Table \ref{mock-samples} we show the combined number of main sequence and giant stars that each of the mentioned surveys will most likely observe within their magnitude-limited volume, distinguished by the birth radius from which they have escaped. We have normalized these values for each of the 4 zones from where we took our samples in our simulations as mentioned before. For the GALAH-AAO samples we have first considered all the stars that this survey will observe and then the number of stars observed that might belong to the same original cluster, i.e. we have normalized the total number of stars by the number of clusters that we have modelled at each radius.

Since we have considered only a discrete sample of cluster masses and birth locations our results should be scaled in some appropriate manner, for example by utilizing a cluster initial mass function (CIMF).

For now we will simply conduct a rough order of magnitude estimate based on the mass of our sample and the mass of the galaxy, but we note that our cluster masses are within the most likely range of the CIMF of \citet{Piskunov}, noting also that the CIMF cluster masses contain a fair degree of uncertainty owing to the difficulty of establishing tidal radii for sparse clusters in dense backgrounds.

We have represented the entire stellar component of the Milky Way with a combined $4.3 \times 10^5 \, {\rm M}_\odot$ of stars born in our sampled clusters. The total stellar mass of the Milky Way is $\sim 3 \times 10^{10} \, {\rm M}_\odot$
\citep{Flynn2006}, which means that the factor needed to convert both, the stellar surface density around the Sun and the total number of stars from our simulations to the real Galaxy is $\sim 1 \times 10^5$
-- assuming all stars are born in bound star clusters, which is an overestimate by at least a factor of two.

We should also add that so far we have assumed that all of our clusters were born at the same time whereas if we instead assumed random birth times that would introduce an additional factor of two correction.
However these approximations can be easily absorbed into the current order of magnitude
exercise.

After combining the factor of $\sim 1 \times 10^5$ with the total results for the three surveys analysed in Table 5 we reach the following values: $\sim 3 \times 10^8$ stars for {\it{Gaia}} (which is the expected goal for the survey), $\sim 1.5 \times 10^8$ for faint stars in {\it{Gaia}}-ESO, and an order of $10^5$ stars for GALAH-AAO (which was its initial goal). Finally, we found that with our results
the GALAH-AAO survey will have a 4 per cent chance of finding stars born
from the same $400\, {\rm M}_\odot$ cluster, a roughly 5 per cent chance of finding
stars born from the same $1\,000\, {\rm M}_\odot$ cluster and a 50 per cent chance of
finding stars born from the same $15\,000\, {\rm M}_\odot$ cluster.

\begin{table}
 \caption{Total number of stars in our magnitude-limited sample separated by mass of the birth cluster and Galactocentric birth radius. $Radius$ indicates the birth radius of the parent cluster, {\it{Gaia}} is the total number of stars that the ESA mission will observe with $V \leq$ 20, {\it{Gaia}}-ESO is the total number of stars that the follow-up survey of {\it{Gaia}} will observe with $V \leq$ 19 for faint stars and in parenthesis $V \leq$ 16.5, GALAH is the number of stars that the AAO will observe within a magnitude range between 12 $\leq V \leq 14$, an extinction $E(B-V)$ < 0.2 and with Galactic latitudes above $|b| \leq$ 5 degrees. The final column represent the expected number of stars from the GALAH-AAO sample that will belong to the same original stellar cluster.}
 \label{mock-samples}
 \begin{tabular}{@{}ccccl}
  \hline
  Radius [kpc] & {\it{Gaia}} & {\it{Gaia}}-ESO & GALAH & GALAH$_{cl}$ \\
  \hline
     400 M$_{\odot}$ & & & &\\
  \hline 
   4 & 150 & \,\,\,83 (15) & 0.00 & 0\\
   6 & 194 & 120 (29) & 0.50 & 0.01\\
   8 & 188 & 120 (31)    & 1.00   & 0.03\\   
  10 & 145 & \,\,\,86 (19)    & 0.50 & 0.01\\
  \hline
   1\,000 M$_{\odot}$ & & & &\\
  \hline
   4 & 265 & 156 (26) & 0.25 & 0.01\\
   6 & 374 & 223 (47) & 0.50 & 0.02\\
   8 & 280 & 174 (42) & 1.00 & 0.04\\   
  10 & 244 & 149 (35) & 0.50 & 0.02\\
  \hline
   15\,000 M$_{\odot}$ & & &\\
   \hline
   4 & 214 & 113 (16) & 0.75 & 0.38\\
   6 & 237 & 143 (26) & 1.00 & 0.5\\
   8 & 267 & 168 (41) & 1.00 & 0.5 \\   
  10 & 115 & \,\,\,68 (14) & 0.25 & 0.13\\
  12 & \,\,\,80 & \,\,\,45 (07.25) & 0.50 & 0.25\\
  \hline
  {\bf{Total}} & {\bf{2756}} & {\bf{1654 (353)}} &  {\bf{7.75}} & {\bf{1.88}} \\
    \hline
 \end{tabular}
\end{table}

\section{SUMMARY}
\label{SUMMARY}

Regardless of whether the next big surveys can find any Solar siblings in kinematic or chemical space, they will provide a picture without precedent of the Milky Way. This picture will help us to improve our dynamical and chemical enrichment models of the Galaxy.
To rephrase that, in 
order to unravel the internal evolution of our Galaxy it is imperative that we improve our models of the evolution of the Milky Way. {\it{Gaia}} will provide in the next years the best kinematic information of stars within 20 kpc which will allow us to better understand the heating mechanisms present in the Galactic disc. 

Two steps in order were taken on this work: 1) develop a Galactic model that closely matches observations, particularly the latest determination for mass and scalelength of the Galactic disc \citep{Bovy} where we especially wanted to match the heating radial profile and history of the Galactic disc, and 2) tracking the continued dynamical evolution of stars that escaped from stellar clusters that were evolved with NBODY6.

We focussed our efforts on analysing the state of the escapers within the Solar annulus after 4.5 Gyr of evolution. In our study we have assumed that all our clusters were born at the same time, i.e. 4.5 Gyr ago. The main analysis included the number of binary systems that ended in the Solar annulus after 4.5 Gyr of evolution. This value is roughly the same fraction as the total contribution of all the modelled clusters, i.e. 30 per cent. We studied the number of Solar siblings and Solar twins that ended in the Solar neighbourhood after 4.5 Gyr.

We determine the extent to which the dynamical signatures of our modelled clusters could be still identified. In the case of our 400 M$_{\odot}$ clusters, overdensities with respect to the Galactic sea of stars were erased after 500 Myr past dissolution. The time-scale of memory lost for overdensities for our 1\,000 M$_{\odot}$ clusters is greater than twice the value for the 400 M$_{\odot}$ clusters, which relates to the different dissolution time-scales of these two sets.

Finally, we study the possibility that the Sun could have originated from any of the three different cluster families that we have modelled. Even when we can claim that the Sun originated in a cluster ranging from  400 -- 1\,000 M$_{\odot}$ we were aware that there exists a lot of evidence that suggests that the potential parent cluster of the Sun could be as high as 15\,000 M$_{\odot}$ or more (\citealt{Adams}; \citealt{Joss2010}). Having this in mind we have included a set of star cluster models with 15\,000 M$_{\odot}$ and we analysed as well the probability of the Sun to be born in one of them.

We provided estimates of the number of stars that three upcoming surveys ({\it{Gaia}}, {\it{Gaia}}-ESO and GALAH-AAO) will observe from clusters like the ones modelled in this work and particularly the number of stars that might have been born in the same original stellar cluster.

Analysing the chemical similarities between stars formed from the same material, i.e. molecular cloud, is the essential next step in the reconstruction of the building blocks of the Galaxy (\citet{Freeman} and references therein). Most chemodynamical codes like RAMSES-CH \citep{Few}, GCD+ \citep{Kawata} and semi-numerical models (e.g. \citealt{Fenner2003}; \citealt{Matteucci2004}) have been successful in explaining the coarse-evolution of the metals over cosmological time. Including nucleosynthesis information on small scales, i.e. stars (Moyano Loyola \& Hurley, in prep.) will improve the fine details in which the chemical evolution of the Galaxy is studied. For this we are making efforts to implement a modular stellar evolution approach in {\tt{NBODY6}} that includes nucleosynthesis information. On the observational side, AAO-HERMES, ESO-WEAVE \citep{Dalton2012}, and 4MOST \citep{deJong} are the primary instruments that will allow the identification of the chemical signatures of stars that have been stripped from a given star cluster beyond the point where the kinematics has been lost.

\section*{ACKNOWLEDGEMENTS}

This work was performed on the gSTAR national facility at Swinburne University of Technology. gSTAR is funded by Swinburne and the Australian Government's Education Investment Fund. GML thanks Joss Bland-Hawthorn for helpful suggestions that improved this work. GML wants also to thank Swinburne University of Technology for the SUPRA scholarship and the Australian Astronomical Society (ASA) for granting financial support via the ASA Travel Assistance scheme. BKG acknowledges the support of the UKs Science \& Technology Facilities Council (ST/J001341/1). The authors thank Ben Macfarlane for his input during the design of TraCD. The authors thank the referee Alex Rastorguev for his many suggestions that helped improve this work.

\bibliography{biblio}
\bibliographystyle{mn2e}



\label{lastpage}

\end{document}